\documentclass{emulateapj}

\newcommand\hour{\mbox{$^{\mathrm h}$}}
\newcommand{\hmpc}{\,h^{-1}\,{\rm Mpc} }
\newcommand{\kpc}{\, {\rm kpc} }
\newcommand{\mpc}{\,{\rm Mpc} }
\newcommand{\msun}{\,{\rm M}_\odot }

\def\eg{{e.g., }}
\def\ie{{i.e., }}
\def\spose#1{\hbox to 0pt{#1\hss}}
\newcommand{\half}{{\textstyle{1\over2}} }
\newcommand{\lta}{\lesssim}
\newcommand{\gta}{\gtrsim}

\newcommand{\COBE}{{\it COBE}}
\newcommand{\DMR}{DMR}
\newcommand{\TOCO}{TOCO}
\newcommand{\Boomerang}{BOOMERANG}
\newcommand{\Maxima}{MAXIMA}
\newcommand{\DASI}{DASI}
\newcommand{\VSA} {VSA}
\newcommand{\mCBIofg}{\textsc{CBI}o140}
\newcommand{\mCBIefg}{\textsc{CBI}e140}
\newcommand{\mCBIocg}{\textsc{CBI}o200}
\newcommand{\mCBIecg}{\textsc{CBI}e200}
\newcommand{\dCBI}{\textsc{CBI}deep}
\newcommand{\cbi}{CBI}

\begin{document}

\journalinfo{Accepted for publication in The Astrophysical Journal}
\submitted{}

\title{Cosmological Parameters from Cosmic Background Imager
  Observations  and Comparisons with BOOMERANG, DASI, and MAXIMA}

\shortauthors{SIEVERS ET AL.}
\shorttitle{COSMOLOGICAL PARAMETERS FROM THE COSMIC BACKGROUND IMAGER}

\author{J.~L.~Sievers,\altaffilmark{1}
J.~R.~Bond,\altaffilmark{2}
J.~K.~Cartwright,\altaffilmark{1}
C.~R.~Contaldi,\altaffilmark{2}
B.~S.~Mason,\altaffilmark{1}
S.~T.~Myers,\altaffilmark{3}
S.~Padin,\altaffilmark{1}
T.~J.~Pearson,\altaffilmark{1}
U.-L.~Pen,\altaffilmark{2}
D.~Pogosyan,\altaffilmark{2,4}
S.~Prunet,\altaffilmark{2,5}
A.~C.~S.~Readhead,\altaffilmark{1}
M.~C.~Shepherd,\altaffilmark{1}
P.~S.~Udomprasert,\altaffilmark{1}
L.~Bronfman,\altaffilmark{6}
W.~L.~Holzapfel,\altaffilmark{7}
and
J.~May\altaffilmark{6}}

\altaffiltext{1}{Owens Valley Radio Observatory, California Institute of
Technology, 1200 East California Boulevard, Pasadena, CA 91125}

\altaffiltext{2}{Canadian Institute for Theoretical Astrophysics, 60 St. George
Street, Toronto Ontario M5S 3H8}

\altaffiltext{3}{National Radio Astronomy Observatory, P.O. Box O, Socorro, NM 87801}

\altaffiltext{4}{Physics Department, University of Alberta, Edmonton, Canada}

\altaffiltext{5}{Institut d'Astrophysique de Paris, 98bis Boulevard Arago, F 75014 Paris, France}

\altaffiltext{6}{Departamento de Astronom\'{\i}a, Universidad de Chile,
Casilla 36-D, Santiago, Chile}

\altaffiltext{7}{University of California, 426 LeConte Hall, Berkeley, CA 94720-7300}

\begin{abstract}
We report on the cosmological parameters derived from observations
with the Cosmic Background Imager (\cbi), covering 40 square degrees
and the multipole range $300 \lesssim \ell \lesssim 3500$.  The angular scales
probed by the \cbi\ correspond to structures which cover the mass
range from 
$10^{14} M_\sun$ to $10^{17} M_\sun$, and the
observations reveal, for the first time, the seeds that gave rise to
clusters of galaxies.  These unique, high-resolution observations
also show damping in the power spectrum to $\ell \sim 2000$, which we
interpret as due to the finite width of the photon-baryon decoupling
region and the viscosity operating at decoupling.
Because the observations extend to much higher $\ell$
the CBI results provide information complementary to that probed by
the \Boomerang, \DASI, \Maxima, and \VSA\ experiments.  When the \cbi\
observations are used in combination with those from \COBE-\DMR\, we
find evidence for a flat universe, $\Omega_{\rm tot}=1.00_{-0.12}^{+0.11}$
(1-$\sigma$), a power law index of primordial fluctuations, $n_s
=1.08_{-0.10}^{+0.11}$, and densities in cold dark matter,
$\Omega_{\rm cdm}h^2 =0.16_{-0.07}^{+0.08}$, and baryons, $\Omega_{b}h^2
=0.023_{-0.010}^{+0.016}$.  With the addition of large scale structure
priors the $\Omega_{\rm cdm}h^2$ value is sharpened to
$0.10_{-0.03}^{+0.04}$, and we find
$\Omega_{\Lambda}=0.67_{-0.13}^{+0.10}$.  In the $\ell < 1000$ overlap
region with the \Boomerang, \DASI, \Maxima, and \VSA\ experiments, the
agreement between these four experiments is excellent, and we
construct optimal power spectra in the \cbi\ bands which demonstrate
this agreement.  We derive cosmological parameters for the combined
CMB experiments and show that these parameter determinations are
stable as we progress from the weak priors using only CMB observations
and very broad restrictions on cosmic parameters, through the addition
of information from large scale structure surveys, Hubble parameter
determinations and Supernova-1a results.  The combination of these
with CMB observations gives a vacuum energy estimate of
$\Omega_\Lambda = 0.70_{-0.05}^{+0.05}$, a Hubble parameter $h= 0.69
\pm 0.04$ and a cosmological age of $13.7 \pm 0.2$ Gyr. As the
observations are pushed to higher multipoles no anomalies relative to
standard models appear, and extremely good consistency is found
between the cosmological parameters derived for the \cbi\ observations
over the range $610<\ell<2000$ and observations at lower $\ell$.
\end{abstract}

\keywords{cosmic microwave background --- cosmology: observations}

\section{Introduction}\label{sec:intro}

The angular power spectrum of the Cosmic Microwave Background (CMB)
has emerged as a major arena in which our cosmological models and
theories of cosmic structure formation can be tested. In this paper we
estimate cosmological parameters from observed power spectra from the
Cosmic Background Imager (\cbi) and the low $\ell$ anchor of the
\COBE-\DMR\ observations, and we relate the \cbi\ observations to
\Boomerang, \DASI, \Maxima, and earlier CMB experiments. We have also
incorporated results from the \VSA\ \citep{scott02} which were reported at the same
time as these \cbi\ results. We also use
results from large scale structure studies (LSS), supernova
observations (SN1a), and Hubble constant (HST-$h$) measurements to
refine estimates of key cosmological parameters, both for \cbi+\DMR\
alone, and for the \cbi\ results in combination with other CMB observations.

During the year 2000 observing season, the \cbi\ covered three deep
fields of diameter roughly $1^\circ$ \citep[hereafter
Paper~II]{Mason02}, and three mosaic regions, each of size roughly 13
square degrees \citep[hereafter Paper III]{Pearson02}. Methods used
for power spectrum estimation from these interferometry observations
are described in \citet[hereafter Paper IV]{Myers02}.  The results
from the 2001 observing season, when combined with the observations
from 2000, will extend the mosaics to 80 square degrees, roughly
doubling the amount of mosaic data.  This will improve upon the
parameter estimates given here but has yet to be analyzed.

\begin{figure}
\plotone{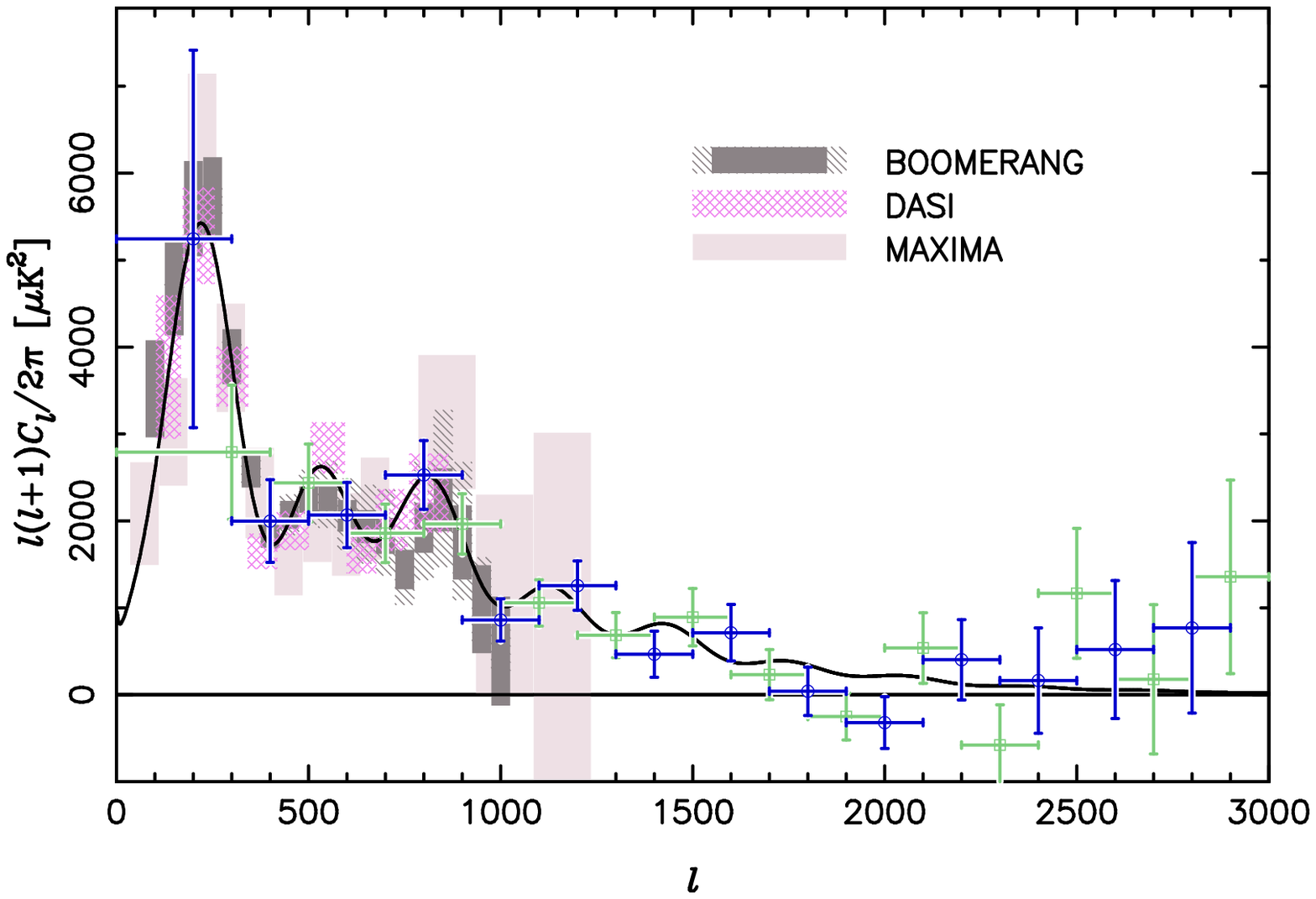}
\caption{Features in the anisotropy spectrum (from Paper~III).  The
first acoustic peak is seen at high sensitivity in the BOOMERANG
\citep{Netterfield02}, DASI \citep{Halverson02}, and MAXIMA
\citep{Lee01} observations, while the second and third acoustic peaks
are seen at lower sensitivity (the rectangles indicate the 68\%
confidence intervals on band-power).  The circles (dark blue) and squares
(green) show the odd and even binnings of the CBI results from the
joint spectrum of the three mosaic fields (see Paper III.  Note that
the two binnings are highly correlated with each other and are not
independent measurements of the power spectrum).  The damping 
tail is clearly seen in the CBI spectrum, and, in the region of overlap,
all four experiments are in excellent agreement, as is discussed in
\S~\ref{sec:optimal2}.  The black curve is the joint model also
discussed in \S~\ref{sec:optimal2}.}
\label{fig:compare}
\end{figure}

Over the past several years, as the CMB observations have improved, the basic
${\cal C}_\ell$ features predicted in the eighties for
inflation-motivated models in which cosmic structure arises from
Gaussian-distributed curvature fluctuations (e.g., Bond \& Efstathiou
1987) have emerged: (1) A ``Sachs-Wolfe'' plateau at low multipole
moments, seen by \COBE-\DMR\ \citep{DMR} and other large angle
experiments. This probes the gravitational potentials at the last
scattering surface and along the line of sight on scales that were not
in causal contact at the epoch of photon decoupling at redshift $z
\sim 1100$.  (2) The long-sought first acoustic peak at $\ell \sim
200$, tentatively first seen by combining results from a heterogeneous
mix of CMB results \citep[e.g.,][]{BJK2000}, then in single
experiments, \TOCO\ \citep{miller00} and \Boomerang-NA North American
Test Flight \citep{mauskopf00}. This was followed shortly after by the
spectacularly detailed first peak mapping by the \Boomerang\ Antarctic
flight \citep{debernardis00,Netterfield02} and \Maxima\ 
\citep{Hanany00,Lee01}; (3) and the detection of the next few peaks
and dips by \Boomerang\ \citep{Netterfield02,debernardis01} and \DASI\ 
\citep{Halverson02}.  Observations from the more recent
experiments are shown in Figure~\ref{fig:compare}. Following these
successes, three other key ingredients remained to be demonstrated
(see, e.g., \citealt{bh95} for a review): (4) The peaks and dips continue
to higher $\ell$ at ever diminishing amplitude, in a damping tail
directly tied to the viscosity in the photon-baryon fluid as they
decouple, and to the finite width of that decoupling region. (5) A
necessary and highly predictable linear polarization power spectrum,
fed by the polarization-dependent Thomson scattering of the primary
CMB anisotropy developed during photon break-out from the decoupling
region. (6) Secondary anisotropies that are an inevitable consequence
as waves develop nonlinearly, breaking to form collapsed structures.

The \cbi\ power spectra show clear evidence for a decline consistent
with the damping tail (4), as can be seen  in
Figure~\ref{fig:compare}.  The Sunyaev-Zeldovich effect has of course
been observed in clusters of galaxies at very high sensitivities, so
(6) is there, and there may be evidence of a statistical SZE signal
from distant clusters in the CBI deep field observations (Papers II and VI),
but further experimental verification is needed.  We address some
issues associated with (4) and (6) in this paper, and (6) is treated
in more detail in \citet[hereafter Paper~VI]{Bond02}. There are many
experiments underway to address (5) and the first detection of
polarization
in the microwave background has recently been announced 
\citet{leitch02,kovac02}; \cbi\ has been observing polarization since
September 2002. A longer
term possibility would provide a 7th pillar, the detection of a
component attributable to gravity waves in the CMB observations. However,
unlike the other six items, inflation models differ substantially in the
amount of gravity waves predicted, with very small undetectable
imprints on the CMB being quite feasible.

One reason that theory can be so definitive is that the primary CMB
anisotropies probe the linear regime of fluctuations and can be
calculated in exquisite detail. The
adiabatic inflation-motivated paradigm continues to do remarkably well
as our knowledge of ${\cal C}_\ell$ and other parameters improves. Just as
powerful though is the large reduction in the space of possible
theories which accurate measurements of the anisotropies provide. The
positioning of the peaks is a strong argument in favor of
predominantly curvature (adiabatic) fluctuations as opposed to
isocurvature ones. In addition the multiple peaks are a strong
argument in favour of coherent (passive) perturbations as opposed to
incoherent (active) ones, e.g., those associated with cosmic defect
theories of structure formation 
\citep{Allen97,Turok98,Contaldi99}.

Minimal inflation-based models characterize the predictions with a
handful of cosmological parameters $\{\Omega_{\rm {tot}}$,
$\Omega_\Lambda$, $\Omega_b h^2$, $\Omega_{\rm cdm} h^2$, $n_s$, $\tau_C$,
$\ln {\cal C}_{10}\}$. These are now very familiar to astrophysicists
\citep[see, e.g.,][]{bet,lange01,jaffe00,Pryke2002,Netterfield02}. The
present-day density $\rho_j$ of a component $j$ is $\Omega_j = 8\pi
G\rho_j/(3H_0^2)$ where $H_0=100h\;$km~s$^{-1}$~Mpc$^{-1}$ is the Hubble
constant. We use the notation $\omega_j \equiv \Omega_j h^2$ for the
related physical density, which is more relevant for the CMB. We
therefore consider the densities of baryons, $\omega_b$, and of cold dark
matter (CDM), $\omega_{\rm cdm}$, with total matter density parameter
$\Omega_m=\Omega_{\rm cdm}+\Omega_b$. The energy density parameter associated
with a cosmological constant is $\Omega_\Lambda$ and the total density
is $\Omega_{\rm {tot}}=\Omega_m+\Omega_\Lambda$, related to the curvature
energy density parameter by $\Omega_k = 1- \Omega_{\rm {tot}}$.  The
initial spectrum of density perturbations is described by an
amplitude, which we usually take to be a factor $\mathcal{C}_{10}$
multiplying the CMB spectrum, and the spectral tilt of scalar
(density) perturbations, $n_s$ (defined so that the initial
three-dimensional perturbation power in gravitational potential
fluctuations per $\ln k$ is $\propto k^{n_s-1}$, with $n_s=1$ thus
giving a scale invariant spectrum).  We also consider the Thomson
depth $\tau_C$ to the epoch of reionization, presumably associated
with luminous star formation in the earliest forming dwarf galaxies.
Many more parameters may be needed to completely describe inflationary
models. These include the gravity-wave induced tensor
amplitude and tilt, variations of tilt with wavenumber, relativistic
particle densities, more complex dynamics associated with the dark
energy $\Omega_\Lambda$, etc. We discuss these briefly below, but
concentrate on the minimal set of 7 parameters for this paper. The
grid of parameter values we have used in this work is an extended
version of the one used
by \citet{lange01}, and is reproduced in Table~\ref{tab:grid}.

\begin{deluxetable*}{llllllllllllll}
\tabletypesize{\scriptsize}
\tablecaption{Parameter Grid for Likelihood Analysis
\label{tab:grid}}
\tablewidth{0pt}
\tablehead{\colhead{Parameter}
&Grid:}
\startdata
$\Omega_k$
& 0.9& 0.7& 0.5& 0.3& 0.2& 0.15& 0.1& 0.05&
0& -0.05& -0.1&-0.15
\\
$\;$& -0.2& -0.3& -0.5
\\
$\omega_{\rm cdm}$& 0.03& 0.06& 0.08 & 0.10 & 0.12& 0.14& 0.17& 0.22& 0.27& 0.33& 0.40& 0.55& 0.8
\\
$\omega_b $& 0.003125& 0.00625& 0.0125& 0.0175& 0.020& 0.0225& 0.025& 0.030&
0.035& 0.04& 0.05& 0.075
\\
$\;$& 0.10& 0.15& 0.2
\\
$\Omega_\Lambda $& 0& 0.1& 0.2& 0.3& 0.4& 0.5& 0.6& 0.7& 0.8& 0.9& 1.0&
1.1
\\
$n_s  $& 1.5& 1.45& 1.4& 1.35& 1.3& 1.25& 1.2& 1.175&
1.15& 1.125& 1.1
\\
$\;$& 1.075& 1.05& 1.025& 1.0
& 0.975& 0.95& 0.925& 0.9& 0.875& 0.85& 0.825
\\
& 0.8&
0.775& 0.75& 0.725& 0.7& 0.65& 0.6& 0.55& 0.5
\\
$\tau_c $& 0& 0.025& 0.05& 0.075& 0.1& 0.15& 0.2& 0.3& 0.4& 0.5& 0.7
\\
\enddata
\end{deluxetable*}

The main target of the CBI, as with most other experiments to date, is
to measure primary anisotropies of the CMB, those which can be
calculated using linear perturbation theory. The maps shown in Papers
III and VI are, to a first approximation, images of damped sound wave
patterns that existed about 400,000 years after the Big Bang at a time
when the photons were freed from the plasma. However, the images are
actually a projected mixture of dominant and subdominant physical
processes occurring through the photon decoupling ``surface'', a fuzzy
wall at redshift $z_{\rm dec} \sim 1050$, when the Universe passed from
optically thick to thin to Thomson scattering over a comoving distance
$\sim 7 \, \omega_m^{-1/2} \mpc$. (Specific numbers here are
appropriate for a $\Lambda$CDM universe preferred by the parameter
estimates in this paper; $\omega_m \sim 0.14$ is one of our
conclusions.)  Prior to this epoch, acoustic wave patterns in the
tightly-coupled photon-baryon fluid on scales below the comoving
``sound crossing distance'' at decoupling, $\sim 50\, \omega_m^{-1/2}
\mpc$ (\ie $\sim 50\, \omega_m^{-1/2} \kpc$ physical), were viscously
damped and strongly so on scales below the damping scale $\sim 4\,
\omega_m^{-1/2}\mpc$. This is closely related to the thickness over
which decoupling occurred. Subsequently, the photons freely-streamed
along geodesics to us, mapping (through the angular diameter distance
relation) the post-decoupling spatial structures in the temperature to
the angular patterns we observe now as the primary CMB
anisotropies. For example, the sound crossing and damping scales
translate to multipoles $\sim 110$ and $\sim 1300$,
respectively. Free-streaming along our (linearly perturbed) past light
cone leaves the pattern largely unaffected, except for the effect of
temporal evolution in the gravitational potential wells as the photons
propagate through them which leaves a further $\Delta T$ imprint,
known as the integrated Sachs-Wolfe effect.

A number of effects complicate this simple picture of direct mapping
of acoustic compression and rarefaction regions: anisotropies are also
fed by the electron flow at decoupling leading to a Doppler effect,
the changing gravitational potential as the universe passes from
domination by relativistic to nonrelativistic species and small terms
associated with polarization. The damping is of course a
major radiative transfer problem, connecting the tightly-coupled 
baryon-photon fluid regime when shear viscosity and thermal conduction
accurately describe the damping through to the free-streaming regime
when full transport is needed. Intense theoretical work over three
decades has put accurate calculations of this linear cosmological
radiative transfer problem on a firm footing. We discuss this further
in \S~\ref{sec:damp}.

Of course there are a number of nonlinear effects that are also
present in the maps. These {\it secondary} anisotropies include
weak-lensing by intervening mass, Thomson-scattering by the nonlinear
flowing gas once it became ``reionized'' at $z \sim 10$--$20$, the thermal and
kinematic SZ effects, and the red-shifted emission from dusty
galaxies. They all leave non-Gaussian imprints on the CMB
sky. Theoretical predictions based on the best-fit models suggest that
for CBI, only the thermal SZ effect would be within striking
distance. This effect is addressed in Paper~VI.

The structure of the paper is as follows. In \S~\ref{sec:priors} we
first summarize the various priors from non-CMB observations that we use 
when estimating cosmological parameters. In \S~\ref{sec:ps} and 
Appendix~\ref{app:ps}, we describe the methods we have 
used for parameter determination from power spectrum estimates, and
the tests we have performed to demonstrate accuracy. 
We obtain estimates for our minimal inflation-based parameter set in
\S~\ref{sec:cbionly}, and also apply the techniques of Appendix~\ref{app:ps}
to determine optimal bandpower spectra in \S~\ref{sec:optimal1}.  We
check for consistency between the mosaic and deep field results
by comparing optimally calculated spectra for various subsets of the
data.  In the parameter estimations we restrict ourselves to the $\ell
< 2000$ region. The spectrum at $\ell > 2000$ is discussed in Paper~VI
together with the level of secondary anisotropy expected from the
Sunyaev-Zeldovich effect in that regime, and whether this is the
origin of the excess power seen in the \cbi\ observations at high
$\ell$.  In \S~\ref{sec:optimal1} we compare our results with those of
other CMB experiments as a further consistency check. We then combine
all available results with the \cbi\ observations to obtain an all
inclusive set of parameter estimates in \S~\ref{sec:allparams}. 
In \S~\ref{sec:damp} we summarize the physical effects probed directly by
the CBI observations. 
Our conclusions are presented in \S~\ref{sec:concl}.  

\section{Prior Probabilities Used in Cosmological Parameter Extraction}
\label{sec:priors}

We apply a sequence of increasingly strong ``prior'' probabilities
successively to the likelihood functions.  These are well known from
the CMB literature \citep[e.g.,][]{bj99,lange01,jaffe00,Netterfield02}.
The priors we apply are listed below.

\noindent 1) The ``weak-$h$'' prior: this prior restricts the Hubble
parameter to $0.45 < h < 0.9$, and also imposes an age restriction,
$t_0>10$ Gyr, and a restriction on the matter density, $\Omega_m >
0.1$.  These are all weak priors that most cosmologists would readily
agree on.

\noindent 2) The ``flat'' prior: as we shall see, the \cbi\
observations strongly support $\Omega_k \approx 0$, so the addition of
a flat prior to weak-$h$ seems reasonable. This is especially so if the
target is parameters associated with inflation models. Although it is
possible for inflation models to give large mean curvature with
non-negligible $\vert \Omega_{k}\vert $, they are rather baroque.

\noindent 3) The ``LSS'' prior: The LSS prior we use here is slightly modified
over that used earlier \citep{bj99,lange01,jaffe00,Netterfield02} and is
described in detail in Paper~VI. It involves a constraint on the
amplitude $\sigma_8^2$ and shape $\Gamma_{\rm eff}$ of the (linear)
density power spectrum. Here $\sigma_8$ is the rms density power
on scales corresponding to rich clusters of galaxies ($8\hmpc$) and
$\Gamma_{\rm eff}$ mainly parameterizes the critical length scale when the
universe passed from dominance by relativistic matter to dominance by
non-relativistic matter. Both constraints depend upon our basic
minimal parameter set in complex ways. The distributions in both these
LSS parameters are taken to be quite broad, akin to a ``weak LSS
prior''.  We take a distribution for the combination $\sigma_8
\Omega_m^{0.56}$ which is a Gaussian (first error) smeared by a
uniform (top-hat) distribution (second error):
$0.47^{+.02,+.11}_{-.02,-.08}$. The 0.47 value is about 15\% below the
value adopted in the earlier CMB papers, as discussed in Paper~VI, but
the distribution shape is the same. To use it, the relation between
${\cal C}_{10}$ and $\sigma_8^2$ is needed for each cosmological model
considered. For the shape prior, we make use of the similarity, over
the wavenumber band that most large scale structure data probes,
between changes in the spectral index $n_s$ and changes in the shape
parameter $\Gamma \approx \Omega_m h \,
\exp[-\Omega_B(1+\Omega_m^{-1}(2{ h})^{1/2})]$.  The latter includes a
strong dependence upon $\Omega_m h$ as well as a rough $\Omega_B$
modification.  We combine the two dependences into a single constraint
on the parameter $\Gamma_{\rm eff}$ = $\Gamma + (n_s - 1)/2$ with
$\Gamma_{\rm eff}$ = $0.21^{+.03,+.08}_{-.03,-.08}$, a broad distribution
over the 0.1 to 0.3 range. It is slightly less skewed to lower
$\Gamma_{\rm eff}$ than the distribution used in the earlier studies. 
These small changes, which accord better with the emerging
LSS data from the 2dF \citep{Peacock01}, SDSS \citep{Szalay01} and
weak lensing surveys \citep{Hoekstra02,Ludo02,Refregier02,Bacon02},
have a very small impact on the cosmic parameters we derive when the
LSS prior is applied, mainly because the CMB results are now so good that
the LSS prior is not as powerful a delimiter as it used to be.

\noindent 4) The ``HST-h'' prior: the HST key project has led to more
restrictive estimates of the Hubble parameter using Cepheid data:
$h = 0.72 \pm 0.08$~\citep{mould00,freedman01}, where these are Gaussian 1-sigma
errors.  We denote this in the paper by the HST-$h$ prior.

\noindent 5) The ``SN'' prior: Comparison of observations of a large number of
distant and nearby supernovae of Type 1a leads to a constraint in the
$\Omega_m -\Omega_\Lambda$ plane, or equivalently the $\Omega_k
-\Omega_\Lambda$ plane~\citep{perlmutter,riess}, independent of the other cosmic
parameters in our minimal set.

\section{Parameterized Power Spectra from Radically Compressed Bandpowers}
\label{sec:ps}

We wish to determine likelihood functions ${\cal L}(y^a) = P(D\vert
y^a)$ for data sets $D$ as a function of parameters $y^a$. These can be
cosmological, as in the minimal inflation set described above;
bandpowers for a discrete binning of ${\cal C}_\ell$ spectra; or
experimental, as for calibration and beam uncertainties. We consider
two classes of parameters, those which are constrained at prescribed
values (external) and those which we allow to dynamically relax to their
maximum likelihood values (internal). For the cosmological parameter
set, we treat all but ${\cal C}_{10} $ as external, with $\ln {\cal
L}(y^a)$ determined on a 8-million point grid. Calibrations, beam
uncertainties, bandpowers and ${\cal C}_{10} $ are treated as
internal, with error estimates in the neighbourhood of the maximum
likelihood made from the curvature (second derivative) matrix
evaluated there.  For example, for the optimal bandpower case we treat
in \S~\ref{sec:optimal1}, the parameterization is of the simple form:
\begin{equation}
{\cal C}_\ell = \sum_b y^b
{\cal C}_\ell^{(s)} \psi_{b\ell}
\label{eq:bins}
\end{equation}
in terms of shapes ${\cal C}_\ell^{(s)}$ and window functions
$\psi_{b\ell}$ which define a partition of unity ($\sum_b \psi_{b\ell}
=1$). An obvious choice for $\psi_{b\ell}$ is the top hat
$\chi_{b\ell}$, defined to be unity for $\ell_b \le \ell < \ell_{b+1}$
and zero outside. It is indeed the one we use for optimal spectra.

Ideally we would use all of the information available, e.g., a pixel map
with errors described by a pixel-pixel correlation matrix, to compute
${\cal L}$. This has been done in the past for limited parameter sets,
but the parameter spaces we treat now are large enough that a large
algorithmic speedup would be needed. In practice, we first go through
a stage of ``radically compressing'' the information into a set of
bandpowers with errors. However it is essential for accuracy that the
entire likelihood surface be well represented. We show that this is
true for the \cbi\ observations.  Our data analysis pipeline,
described in Paper IV, grids the visibility data into estimators, the
covariances of which are also computed, and then determines  the
maximum likelihood power spectrum and the curvature of the likelihood
function about that maximum. In the last step, the power spectrum for
a given experiment is parameterized by a discrete sum over contiguous
bands $B$ as in equation (\ref{eq:bins}), with $q^B$ replacing $y^b$ and
$\chi_{B \ell }$ for $\psi_{b\ell}$. For the power spectra estimates
of Papers II and III, a flat shape (${\cal C}^{(s)}_\ell$ constant)
was used.

Even though $\{ \chi_{B \ell } \}$ may be chosen for the input window
functions in the parameterized model of equation (\ref{eq:bins}), processing
through the actual $(u,v)$ coverage results in an effective set of
window functions $\{ W_{B}(\ell ) \}$ (Paper~IV). These are plotted
for the deep and mosaic observations in Papers~II and~III. The $W_{B}(\ell )$
depend upon signal-to-noise  ratio  as a function of
$\ell$ and spill over to $\ell$ values that lie beyond the support of
the top hat $\chi_{B \ell }$.

The result of the ``radical compression'' of the full \cbi\ noisy
visibility data set, as described in Paper IV, is: $\{\bar{q}^B,
{q}_{\rm src}^B, {q}_{\rm res}^B, {q}_{\rm N}^B, (F^{-1})_{BB^\prime},
\varphi_{B\ell} \}$.  Here $\{ \bar{q}^B \}$ are maximum likelihood
values of $\{ {q}^B \}$, $(F^{-1})_{BB^\prime}$ is the inverse
Fisher matrix which would describe the correlations among the
bandpowers if the likelihood distribution were Gaussian, and $\varphi_{B
\ell}$ are the bandpower window functions that convert a spectrum ${\cal
C}_\ell$  to expected bandpower values $\{ \bar{q}^B \}$.
In addition
to the bandpowers, there are often extra parameters associated with
other contributions to the signal that must be simultaneously
determined.  For example, with the \cbi\ data, we have two classes of
sources which we take into account: NVSS sources having flux densities
greater than 3.4 mJy at 1.4 GHz, and ``residual'' faint sources which
we model as an isotropic Gaussian random field. The NVSS sources have
known positions, which enable us to define point-source template
structures on the data. Rather than trying to model the source
amplitudes in detail, we simply project them out of the data by making
the multipliers, ${q}_{\rm src}$, of the templates very large.  We
estimate the amplitude of the residual source contribution by
extrapolating the CBI 31 GHz source counts to fainter flux densities.

We usually fix the uncertainty in those amplitudes as well, but in
some tests have allowed it to vary. In that case, a nuisance parameter
is introduced which increases the Fisher matrix dimension by
one. Forming $(F^{-1})_{BB^\prime}$ and using it in the treatment of
the data is equivalent to having marginalized over the nuisance
parameter, so in effect $(F^{-1})_{BB^\prime}$ is what is needed in
all cases. If we target a limited number of bandpowers for parameter
estimation, $(F^{-1})_{BB^\prime}$ is truncated to those bands. This
is mathematically identical to treating all bandpowers, but
marginalizing (integrating) over the ``unobserved'' bands we have cut
out.

The effective or generalized total noise in each band, ${q}_{\rm Nt}^B
= {q}_{\rm N}^B+{q}_{\rm src}^B+{q}_{\rm res}^B$, includes the noise
itself ${q}_{\rm N}^B$ and the source contributions, ${q}_{\rm
src}^B$ and  ${q}_{\rm res}^B$. Estimation of these
noise and source bandpowers is performed after the maximum likelihood
bandpowers are found, and they are calculated within the same calculational
framework, as described in Paper IV. To calculate theoretical
bandpowers from a given ${\cal C}_\ell$, we also need to specify a set
of window functions relating the contribution of multipole $\ell$ to
band $B$, $\varphi_{B\ell}$. 

The offset lognormal approximation we use to characterize the
likelihood surfaces is described in Appendix~\ref{app:ps} and shown to
be an excellent approximation to our \cbi\ likelihood functions.  

The CBI $\ell$ range and binnings we have used are given in
\S~\ref{sec:cbionly}.  The other experimental results we use have a
variety of $\ell$ coverage and binning: \Boomerang\ covers an $\ell$
range of 75--1125 with band width $\Delta\ell =50$, \DASI\ covers
104--864 with $\Delta\ell$ variable between 70 and 100, \Maxima\
covers 73--1161 with width 75, and \DMR\ covers low $\ell$, from 2 to
30 (although we start from 3 because the quadrupole has a Galactic
contamination). Bands for \TOCO, \Boomerang-NA, and \VSA\ are described by
\citet{miller00}, \citet{mauskopf00}, and \citet{scott02}. The ``Apr99'' combination of
experiments was introduced in \citet{BJK2000}. Together with \cbi\
mosaic, these make up the ``all-data'' combination which we use
extensively when combining CBI with other experiments.

The results that we have used, and the labels by which we refer to
them, are summarized in Table~\ref{tab:datasets}.

\begin{deluxetable*}{ll}
\tabletypesize{\scriptsize}
\tablecaption{Data Sets
\label{tab:datasets}}
\tablewidth{0pt}
\tablehead{\colhead{Label}
& \colhead{Data Set}
}
\startdata
All-data
& Apr99, \Boomerang-NA, \Boomerang, CBIo140, \DASI, \DMR,
\Maxima, \TOCO, \VSA
\\
Apr99
& Compilation of 17 experiments prior to April 1999, by \citet{BJK2000}
\\
\Boomerang-NA 
& North American Test Flight of \Boomerang \citep{mauskopf00}
\\
\Boomerang
& Antarctic
Flight of \Boomerang \citep{debernardis00, debernardis01,
Netterfield02, ruhl02}
\\
\mCBIofg
& CBI 02h+14h+20h mosaics, odd bins, $\Delta \ell=140$
\\
\mCBIocg
& CBI 02h+14h+20h mosaics, odd bins, $\Delta \ell=200$
\\
\mCBIefg
& CBI 02h+14h+20h mosaics, even bins, $\Delta \ell=140$
\\
\mCBIecg 
& CBI 02h+14h+20h mosaics, even bins, $\Delta \ell=200$
\\
\mCBIofg $(\ell>610)$
& CBI 02h+14h+20h mosaics, odd bins, $\Delta \ell=140$, band-powers at
$\ell<610$ discarded
\\
\dCBI
& CBI 08h+14h+20h deep fields
\\
\DASI
& \citep{Halverson02}.
\\
\Maxima
& \citep{Hanany00,Lee01}
\\
TOCO
& \citep{miller00} 
\\
VSA
& \citep{scott02} 
\\
\enddata

\tablecomments{Labels used in the text to designate different data
sets used in the data analysis and data tests }

\end{deluxetable*}

\section{Cosmology with the \cbi\ and Robustness Tests}\label{sec:cbionly}

 In this section we present the cosmological results we
have derived from the CBI observations, and compare these results with those
from non-CMB observations.  In \S~\ref{sec:allparams} we compare the
CBI observations and parameters with those derived from other CMB
experiments and we combine all of these CMB data to determine the best
overall values of the cosmological parameters that can be derived from the CMB
data when combined with large scale structure studies, the HST $H_0$
project, and supernova type 1a results.

 In the derivation of cosmological parameters from the
CBI observations we have carried out a large number of consistency tests and
checks. We find excellent consistency in {\it all} of the tests we
have performed. Some of these tests are described in
\S~\ref{sec:prcbi}, and the more important of these tests are
summarized in \S~\ref{sec:robust}.

\subsection{The Primary CBI Results}\label{sec:prcbi}

The basic set of 7 parameters for our fiducial minimal inflation
model, $\{\Omega_{\rm {tot}}$, $\Omega_\Lambda$, $\Omega_b h^2$,
$\Omega_{\rm cdm} h^2$, $n_s$, $\tau_C$, $\ln {\cal C}_{10}\}$, is described
in the introduction, and the grid of these parameters is given 
 in Table~\ref{tab:grid}.  The amplitude ${\cal
C}_{10}$ is a continuous variable.  The effect on parameter
determinations of the database boundary and the various priors defined
on the space that we use is described in detail by
\citet{lange01}.

As described in Paper III, our standard parameter determinations for
the mosaic observations have been made with bins of width $\Delta \ell =140$,
with two alternate locations of the bins. The ``even'' binning has
$\ell_B = 260 + 140 B$ ($1 \le B \le 23$), while the
``odd'' binning has $\ell_B =  190 + 140 B$ ($1 \le B
\le 23$), where $\ell_B$ is the upper limit of the bin. Data derived
with these binnings are denoted by \mCBIefg\ and \mCBIofg\ in this
paper.  A coarser binning used has width $\Delta \ell =200$, with
``even'' spacing $\ell_B = 200 + 200 B$ ($1 \le B \le 16$) and ``odd''
spacing $\ell_B = 100 + 200 B$ ($1 \le B \le 16$), with the
corresponding data denoted by \mCBIecg\ and \mCBIocg.  Correlations
are strongest between adjacent bins and are typically negative for
interferometry data. The maximum anti-correlation between adjacent
bins is about 25\% for $\Delta\ell = 140$ and about 15\% for
$\Delta\ell=200$.
The \dCBI\ standard bins begin
at 500, 880, 1445, 2010, 2388, 3000, ending at 4000. There is a lower
$\ell$ bin as well, but we do not include it in parameter analysis
(we marginalize over it). 

\begin{figure}
\epsscale{0.9}
\plotone{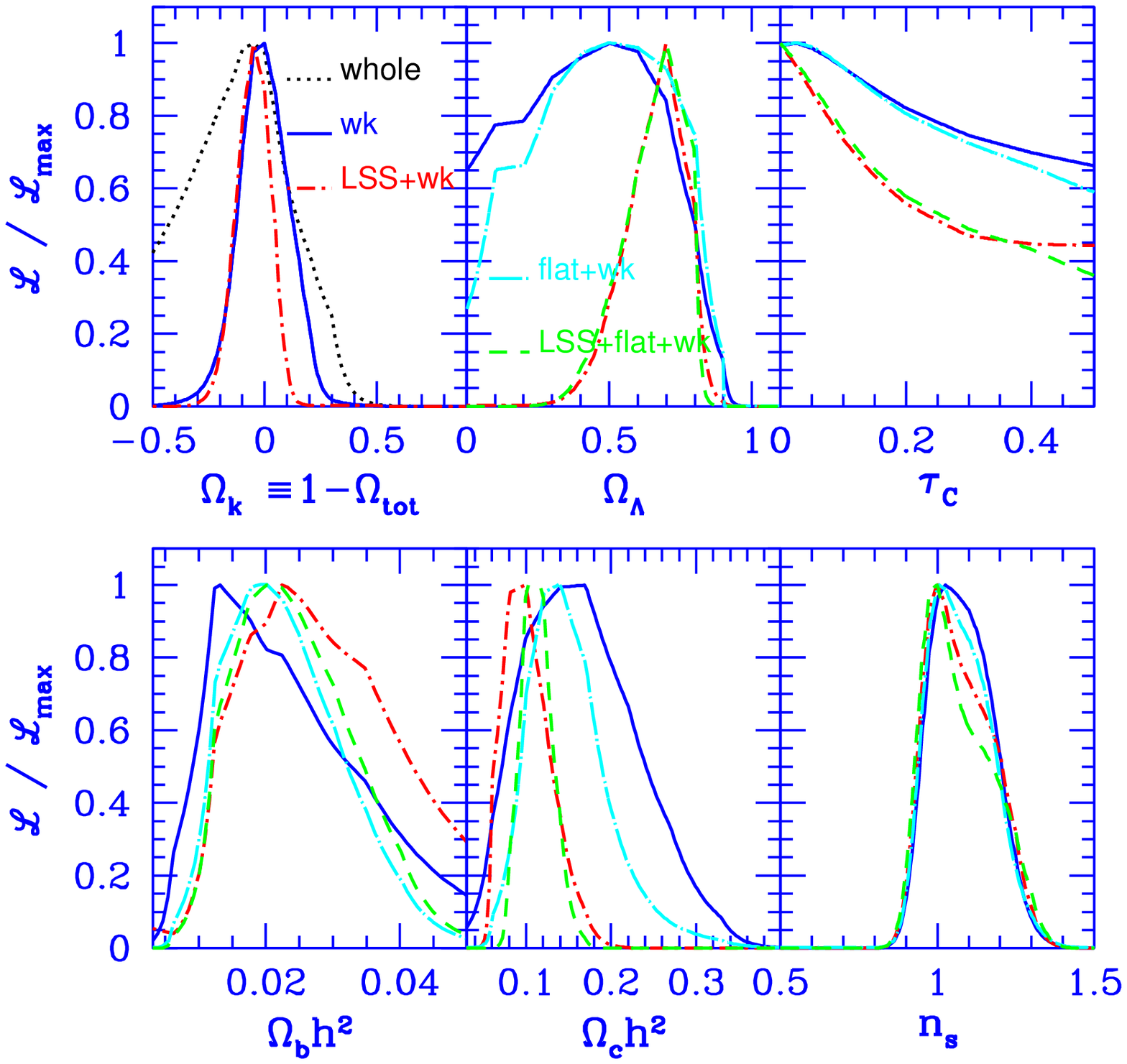}
\caption{1D projected likelihood functions calculated for the
\mCBIofg+\DMR\ data. All panels include the weak-$h$ (solid dark blue) and
LSS+weak-$h$(short-dash-dotted red) priors. (LSS is the large-scale
structure prior.) The $\Omega_k$ panel also shows what the whole
${\cal C}_\ell$-database gives before the weak-$h$ prior is imposed
(black dotted). We note that even in the absence of CMB data there is a
bias towards the closed models~\citep{lange01}. In the other panels,
flat+weak-$h$ (long-dashed-dotted light blue) and LSS+flat+weak-$h$ (dashed
green) are plotted. Notice how stable the $n_s$ determination
is, independent of priors. We see here that, under priors ranging from
the weak-$h$ prior to the weak-$h$+LSS+flat priors, the CBI provides a useful
measure of four out of the six fundamental parameters shown.
 This
is independent of the first acoustic peak, where the CBI has low
sensitivity, and is also largely independent of the spectrum below $\ell
\sim 610$ for all but $\Omega_b h^2$ (see text).}
\label{fig:likecbimO}
\end{figure}

The primary results of the CBI cosmological parameter extraction are
shown in Table~\ref{tab:paramscbiO} and Figure~\ref{fig:likecbimO}.
These show parameter determinations after marginalization over all
other parameters for the \mCBIofg+\DMR\ dataset and for all
combinations of the priors described in \S~\ref{sec:priors}.
 
\begin{deluxetable*}{lllllllllllll}
\tabletypesize{\scriptsize}
\tablecaption{Cosmic Parameters for Various Priors using \mCBIofg+\DMR
\label{tab:paramscbiO}} 
\tablewidth{0pt} 
\tablehead{\colhead{Priors}
& \colhead{$\Omega_{\rm tot}$}
& \colhead{$n_s$}
& \colhead{$\Omega_bh^2$}
& \colhead{$\Omega_{\rm cdm}h^2$}
& \colhead{$\Omega_{\Lambda}$}
& \colhead{$\Omega_m$}
& \colhead{$\Omega_b$}
& \colhead{$h$}
& \colhead{Age}
& \colhead{$\tau_c$}
}
\startdata
wk-$h$
& $1.00^{0.11}_{0.12}$ 
& $1.08^{0.11}_{0.10}$ 
& $0.023^{0.016}_{0.010}$ 
& $0.16^{0.08}_{0.07}$ 
& $0.43^{0.25}_{0.28}$ 
& $0.59^{0.22}_{0.22}$ 
& $0.083^{0.053}_{0.053}$ 
& $0.58^{0.11}_{0.11}$ 
& $13.9^{2.2}_{2.2}$ 
& $<0.66$ 
\\
wk-$h$+LSS
& $1.05^{0.08}_{0.08}$ 
& $1.07^{0.13}_{0.10}$ 
& $0.029^{0.015}_{0.012}$ 
& $0.10^{0.04}_{0.03}$ 
& $0.67^{0.10}_{0.13}$ 
& $0.39^{0.12}_{0.12}$ 
& $0.095^{0.055}_{0.055}$ 
& $0.60^{0.12}_{0.12}$ 
& $15.4^{2.1}_{2.1}$ 
& $<0.66$ 
\\
wk-$h$+SN
& $1.03^{0.08}_{0.08}$ 
& $1.11^{0.11}_{0.11}$ 
& $0.028^{0.016}_{0.012}$ 
& $0.10^{0.05}_{0.04}$ 
& $0.71^{0.08}_{0.09}$ 
& $0.33^{0.08}_{0.08}$ 
& $0.076^{0.046}_{0.046}$ 
& $0.65^{0.12}_{0.12}$ 
& $14.7^{2.4}_{2.4}$ 
& $<0.67$ 
\\
wk-$h$+LSS+SN
& $1.04^{0.08}_{0.08}$ 
& $1.11^{0.11}_{0.10}$ 
& $0.029^{0.016}_{0.012}$ 
& $0.10^{0.04}_{0.03}$ 
& $0.72^{0.07}_{0.07}$ 
& $0.32^{0.08}_{0.08}$ 
& $0.082^{0.047}_{0.047}$ 
& $0.64^{0.11}_{0.11}$ 
& $15.0^{2.2}_{2.2}$ 
& $<0.67$ 
\\
\\[0.01cm]\tableline\\[0.01cm]
flat+wk-$h$
& (1.00) 
& $1.07^{0.11}_{0.10}$ 
& $0.023^{0.010}_{0.008}$ 
& $0.15^{0.06}_{0.04}$ 
& $0.47^{0.25}_{0.27}$ 
& $0.54^{0.24}_{0.24}$ 
& $0.068^{0.028}_{0.028}$ 
& $0.60^{0.12}_{0.12}$ 
& $14.0^{1.4}_{1.4}$ 
& $<0.65$ 
\\
flat+wk-$h$+LSS
& (1.00) 
& $1.05^{0.15}_{0.09}$ 
& $0.024^{0.011}_{0.009}$ 
& $0.11^{0.02}_{0.02}$ 
& $0.67^{0.10}_{0.13}$ 
& $0.34^{0.12}_{0.12}$ 
& $0.057^{0.020}_{0.020}$ 
& $0.66^{0.11}_{0.11}$ 
& $14.2^{1.3}_{1.3}$ 
& $<0.62$ 
\\
flat+wk-$h$+SN
& (1.00) 
& $1.10^{0.12}_{0.11}$ 
& $0.025^{0.011}_{0.009}$ 
& $0.11^{0.03}_{0.02}$ 
& $0.70^{0.07}_{0.07}$ 
& $0.30^{0.07}_{0.07}$ 
& $0.052^{0.017}_{0.017}$ 
& $0.70^{0.09}_{0.09}$ 
& $13.8^{1.4}_{1.4}$ 
& $<0.65$ 
\\
flat+wk-$h$+LSS+SN
& (1.00) 
& $1.08^{0.11}_{0.09}$ 
& $0.025^{0.011}_{0.009}$ 
& $0.11^{0.03}_{0.02}$ 
& $0.71^{0.06}_{0.06}$ 
& $0.29^{0.06}_{0.06}$ 
& $0.053^{0.016}_{0.016}$ 
& $0.69^{0.09}_{0.09}$ 
& $13.9^{1.3}_{1.3}$ 
& $<0.63$ 
\\
\\[0.01cm]\tableline\\[0.01cm]
flat+HST-$h$
& (1.00) 
& $1.09^{0.12}_{0.10}$ 
& $0.026^{0.010}_{0.009}$ 
& $0.13^{0.07}_{0.04}$ 
& $0.65^{0.12}_{0.20}$ 
& $0.38^{0.18}_{0.18}$ 
& $0.058^{0.022}_{0.022}$ 
& $0.68^{0.08}_{0.08}$ 
& $13.3^{1.3}_{1.3}$ 
& $<0.65$ 
\\
flat+HST-$h$+LSS
& (1.00) 
& $1.09^{0.13}_{0.10}$ 
& $0.026^{0.010}_{0.009}$ 
& $0.11^{0.03}_{0.02}$ 
& $0.71^{0.07}_{0.08}$ 
& $0.29^{0.08}_{0.08}$ 
& $0.054^{0.019}_{0.019}$ 
& $0.70^{0.08}_{0.08}$ 
& $13.8^{1.1}_{1.1}$ 
& $<0.64$ 
\\
flat+HST-$h$+SN
& (1.00) 
& $1.10^{0.12}_{0.11}$ 
& $0.026^{0.010}_{0.009}$ 
& $0.12^{0.03}_{0.03}$ 
& $0.71^{0.06}_{0.06}$ 
& $0.29^{0.07}_{0.07}$ 
& $0.052^{0.017}_{0.017}$ 
& $0.71^{0.07}_{0.07}$ 
& $13.6^{1.2}_{1.2}$ 
& $<0.65$ 
\\
flat+HST-$h$+LSS+SN
& (1.00) 
& $1.08^{0.11}_{0.09}$ 
& $0.026^{0.010}_{0.009}$ 
& $0.11^{0.02}_{0.02}$ 
& $0.71^{0.06}_{0.05}$ 
& $0.29^{0.06}_{0.06}$ 
& $0.054^{0.017}_{0.017}$ 
& $0.70^{0.07}_{0.07}$ 
& $13.7^{1.1}_{1.1}$ 
& $<0.63$ 
\\
\enddata 
\tablecomments{Estimates of the 6
external cosmological parameters that characterize our fiducial
minimal-inflation model set as progressively more restrictive prior
probabilities are imposed. ($\tau_C$ is put at the end because it is
relatively poorly constrained, even with the priors.) Central values
and $1\sigma$ limits for the 6 parameters are found from the 16\%,
50\% and 84\% integrals of the marginalized likelihood. For the other
``derived'' parameters listed, the values are means and variances of
the variables calculated over the full probability
distribution. wk-$h$ requires $0.45 < h < 0.90$, Age $>10$~Gyr, and
$\Omega_m>0.1$. The sequence shows what happens when LSS, SN and
LSS+SN priors are imposed. While the first four rows allow
$\Omega_{\rm tot}$ to be free, the next four have $\Omega_{\rm tot}$ pegged to
unity, a number strongly suggested by the CMB data. The final 4 rows
show the ``strong-$h$'' prior, a Gaussian centered on $h=0.71$ with
dispersion $\pm 0.076$, obtained for the Hubble key project. When the
$1\sigma$ errors are large it is usual that there is a poor detection,
and sometimes there can be multiple peaks in the 1D projected
likelihood.   }
\end{deluxetable*}

\subsubsection{The Geometry and the Primordial Fluctuation Spectrum}

We begin the cosmological parameters discussion by considering
$\Omega_{\rm tot}$ and $n_s$.  We see from Table~\ref{tab:paramscbiO} that
under the weak-$h$ prior assumption the combination \mCBIofg+\DMR\
yields $\Omega_{\rm tot}=1.00_{-0.12}^{+0.11}$ and $n_s=1.08_{-0.10}^{+0.11}$.
 
\begin{figure}
\plotone{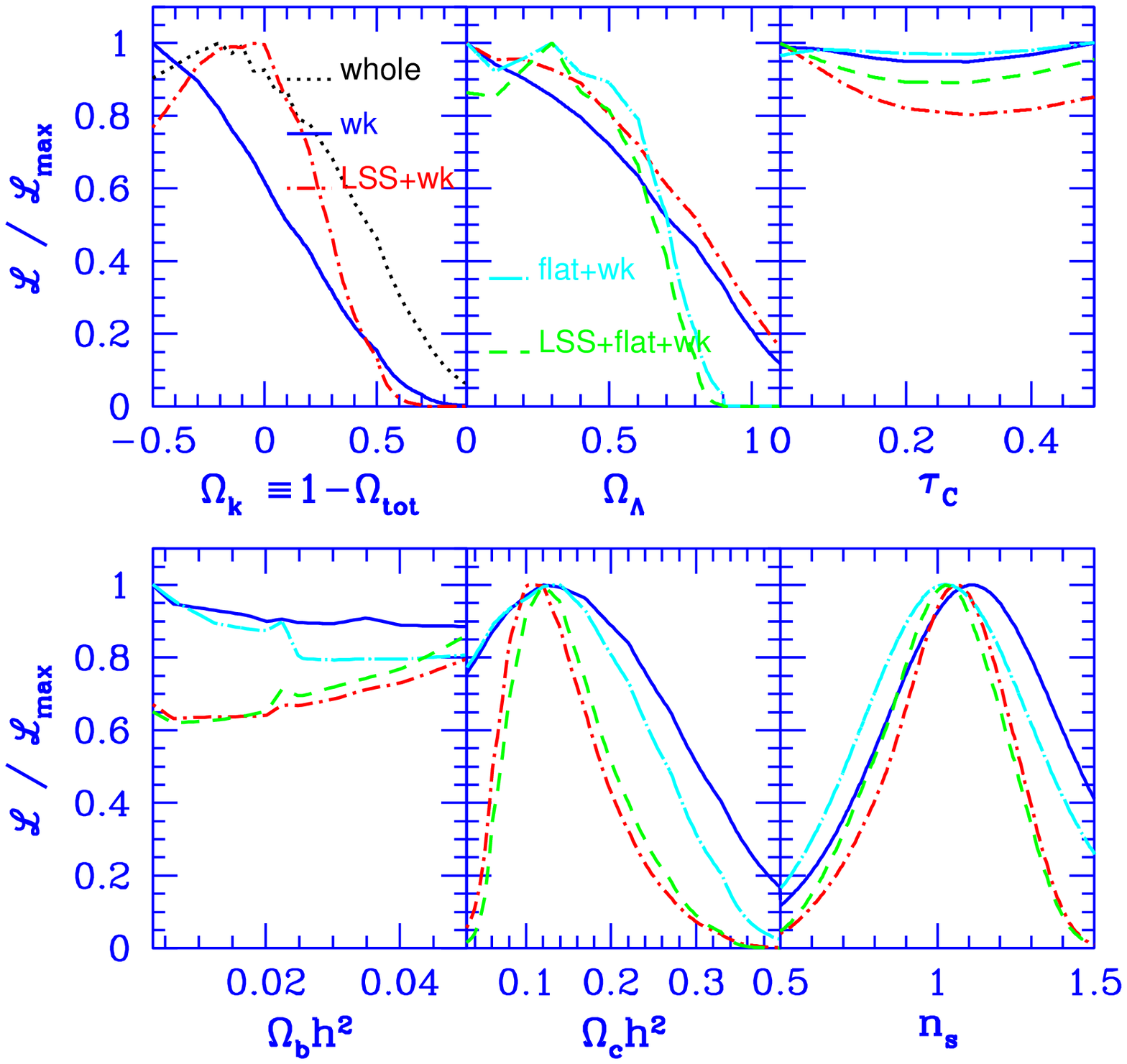}
\caption{Results obtained using \DMR\ alone. This gives an idea of the
role of the LSS prior in sharpening up detections for \DMR. Note that
\DMR\ did reasonably well by itself in first indicating for this class
of models that $n_s \sim 1$ \citep[e.g.,][]{bh95}. Of course it could not
determine $\omega_b$ and the structure in $\Omega_k$ and
$\Omega_\Lambda$ can be traced to ${\cal C}_\ell$-database
constraints~\citep{lange01}. Comparison with Fig.~\ref{fig:likecbimO}
shows the greatly improved constraints when the CBI data are added.}
\label{fig:likedmr}
\end{figure}

To illustrate clearly the constraints that the CBI observations impose
we show in Figure~\ref{fig:likedmr} the likelihoods obtained from DMR
alone under the various priors.  Note that the $\Omega_{\rm k}$
likelihood curve is very broad.  This figure shows that the above
tight cosmological constraints do not arise from the priors.  Rather,
the other experiments nicely complement the CMB, greatly enhancing the
discriminatory power of any single data set.  These likelihood curves
should also be contrasted with the ``prior-only'' likelihood curves
presented by \citet{lange01}.  The results of adding the CBI data are
shown in Figure~\ref{fig:likecbimO}.  The effect of combining the CBI
data with the DMR data under the LSS prior is to reduce the
uncertainties in $\Omega_{\rm tot}$ --- with these priors
$\Omega_{\rm tot}=1.05_{-0.08}^{+0.08}$.  Note that the uncertainties on
$n_s$ are almost independent of the priors, and are always $\sim 10\%$.

\begin{figure}
\plotone{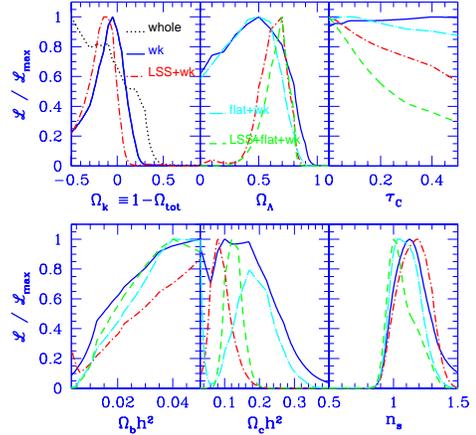}
\caption{Same as Fig.~\ref{fig:likecbimO}, except that CBI data at
  $\ell<610$ have been discarded.  We see that, under the weak-$h$ and
  LSS+weak-$h$  priors, $\Omega_{\rm k}$ is peaked near zero, as is the
  case for the whole data set, showing that in the CBI data the
  evidence for a flat universe is coming from the $\ell$-range 
above the second acoustic peak where
  damping dominates, as well as from
  the lower-$\ell$ range. We see here that under increasingly
  restrictive priors the CBI data at $\ell>610$ provide useful
  constraints on $n_s$, $\Omega_m h^2$ and $\Omega_\Lambda$, showing
  that these results are driven by the shape (and level) of the spectrum at
  high-$\ell$ independent of the results at low-$\ell$.  Note,
  however, that when the CBI data are restricted to $\ell>610$ they do
  not provide a useful measure of $\Omega_b h^2$ (compare with Fig.~\ref{fig:likecbimO}). }
\label{fig:likecbim412}
\end{figure}

The CBI has very
little sensitivity at $\ell<400$, thus these determinations of
$\Omega_{\rm tot}$ and $n_s$ are basically independent of the first
acoustic peak.  We have explored the degree to which the CBI results
depend on the low-$\ell$ data by eliminating the band-powers at $\ell<610$,
and running the same sequence of tests of increasingly restrictive
priors.  The results for the cut at $\ell=610$ are shown in
Figure~\ref{fig:likecbim412} and in Table~\ref{tab:paramscbitests}.
The likelihood is still sharply peaked at $\Omega_{\rm tot}\sim 1$ even
though the data from the region of the first and second peaks has been
discarded.

The effect of discarding the data at $\ell<610$ on $n_s$ is seen in
Figure~\ref{fig:likecbim412}.  We see that the likelihood again peaks
near unity, showing that the determination of a near scale-invariant
fluctuation spectrum is not dependent on the first or second acoustic
peaks in the CBI data.

We conclude that both the geometry and the fluctuation spectrum are
highly consistent with the predictions of the minimal inflationary
theory, and that this consistency applies even when the data at
$\ell<610$, corresponding to the regions of the first and second
acoustic peaks, are discarded.  These results on $\Omega_{\rm tot}$ and
$n_s$ are therefore independent of previous results based on
observations over the multipole range of the first and second acoustic
peaks.
The $\ell<610$ and $\ell>610$ regions of the angular spectrum indicate
a consistent power law for the primordial spectrum, for the minimal
models we consider, over the range of scales covered by the
observations which now extend down to the scales of clusters of galaxies
probed directly by LSS studies.

\subsubsection{The Non-Baryonic and Baryonic Matter Densities and the Cosmological Constant}

The constraints on the density in dark matter, $\omega_{\rm cdm}$, and the
cosmological constant, $\Omega_{\Lambda}$ are shown in
Figure~\ref{fig:likecbimO}.  One sees that these are tightly
constrained when the LSS prior is added. The effect of adding the CBI
data can be seen by comparing the weak-$h$+LSS prior results in
Figure~\ref{fig:likedmr} with those in Figure~\ref{fig:likecbimO}.  We
see that the CBI data reduce the uncertainties dramatically, and that
$\omega_{\rm cdm}=0.10_{-0.03}^{+0.04}$ and
$\Omega_{\Lambda}=0.67_{-0.13}^{+0.10}$ for the weak-$h$+LSS priors case.
If, in addition, we assume a flat geometry, the uncertainties in the
dark matter density are further improved,
$\omega_{\rm cdm}=0.11_{-0.02}^{+0.02}$, while those for the cosmological
constant are the same.  We see from Figure~\ref{fig:likecbim412}
and from Table~\ref{tab:paramscbitests}
that the constraints on the dark matter density are also reasonably tight when
the data at $\ell<610$ are discarded.  The results on
$\Omega_{\Lambda}$ are little changed when the data below $\ell=610$
are discarded although the uncertainties are larger, as can be seen by comparing
Figure~\ref{fig:likecbim412} with Figure~\ref{fig:likecbimO}.  Thus
these results on $\omega_{\rm cdm}$ and $\Omega_\Lambda$ are also
established over the high-$\ell$ range independent of the first and
second acoustic peaks.

The fractional CBI constraints on the baryonic matter density are not
nearly as tight as those on non-baryonic matter.  With the
weak-$h$+LSS+flat priors, we see from Table~\ref{tab:paramscbiO} that
$\omega_b= 0.024_{-0.009}^{+0.011}$.  This is consistent with the
results from Big Bang nucleosynthesis \citep{olive,burles,tytler}.  When
we discard the data below $\ell=610$, we do not yet have a useful
constraint on $\Omega_b h^2$ (see Figure~\ref{fig:likecbim412}).
Although weaker, the potential constraints on the baryonic matter
density from
the high-$\ell$ region of the spectrum are important because this is,
in principle, a direct and independent way of measuring $\Omega_b
h^2$.

From the combination of the densities in both baryonic and
non-baryonic matter, we see that the CBI provides compelling evidence
for a matter density significantly lower than the critical density
required to close the universe, with $\Omega_m=0.34 \pm 0.12$, with 
$\Omega_b=0.057\pm0.020$ in baryons.  This result, combined with the flat
geometry, requires a significant energy component of the universe to
be supplied by something other than matter, and this we assume under
our minimal inflationary scenario to be the cosmological constant.

\subsubsection{The Hubble Constant and the Age of the Universe}

Significant measures of the Hubble constant and the age of the
universe are again obtained under the weak-$h$+LSS+flat priors (see
Table~\ref{tab:paramscbiO}).  We find, under these priors, that
$h=0.66\pm0.11$ and $t_0=14.2 \pm 1.3$ Gyr.  These are in excellent
agreement with recent determinations of the Hubble constant
\citep{mould00,freedman01} and the ages of the oldest stars in globular
clusters.

\subsection{Further Robustness Tests}
\label{sec:robust}

We have carried out a large number of tests, further to those
described in the previous section, of the parameter extraction from
the CBI observations.  We describe  some of the more important of these tests
in this section.

\begin{deluxetable*}{lllllllllll}
\tabletypesize{\scriptsize}
\tablecaption{\cbi\ Tests and Comparisons
\label{tab:paramscbitests}}
\tablewidth{0pt}
\tablehead{\colhead{Priors}
& \colhead{$\Omega_{\rm tot}$}
& \colhead{$n_s$}
& \colhead{$\Omega_bh^2$}
& \colhead{$\Omega_{\rm cdm}h^2$}
& \colhead{$\Omega_{\Lambda}$}
& \colhead{$\Omega_m$}
& \colhead{$\Omega_b$}
& \colhead{$h$}
& \colhead{Age}
& \colhead{$\tau_c$}
}
\startdata
\mCBIofg & & & & & & & & & & \\
wk-$h$
& $1.00^{0.11}_{0.12}$ 
& $1.08^{0.11}_{0.10}$ 
& $0.023^{0.016}_{0.010}$ 
& $0.16^{0.08}_{0.07}$ 
& $0.43^{0.25}_{0.28}$ 
& $0.59^{0.22}_{0.22}$ 
& $0.083^{0.053}_{0.053}$ 
& $0.58^{0.11}_{0.11}$ 
& $13.9^{2.2}_{2.2}$ 
& $<0.66$ 
\\
flat+wk-$h$
& (1.00) 
& $1.07^{0.11}_{0.10}$ 
& $0.023^{0.010}_{0.008}$ 
& $0.15^{0.06}_{0.04}$ 
& $0.47^{0.25}_{0.27}$ 
& $0.54^{0.24}_{0.24}$ 
& $0.068^{0.028}_{0.028}$ 
& $0.60^{0.12}_{0.12}$ 
& $14.0^{1.4}_{1.4}$ 
& $<0.65$ 
\\
flat+wk-$h$+LSS
& (1.00) 
& $1.05^{0.15}_{0.09}$ 
& $0.024^{0.011}_{0.009}$ 
& $0.11^{0.02}_{0.02}$ 
& $0.67^{0.10}_{0.13}$ 
& $0.34^{0.12}_{0.12}$ 
& $0.057^{0.020}_{0.020}$ 
& $0.66^{0.11}_{0.11}$ 
& $14.2^{1.3}_{1.3}$ 
& $<0.62$ 
\\
\\[0.01cm]\tableline\\[0.01cm]
\mCBIefg & & & & & & & & & & \\
wk-$h$
& $0.96^{0.14}_{0.13}$ 
& $1.10^{0.11}_{0.11}$ 
& $0.016^{0.013}_{0.007}$ 
& $0.18^{0.08}_{0.07}$ 
& $0.37^{0.28}_{0.26}$ 
& $0.62^{0.23}_{0.23}$ 
& $0.059^{0.044}_{0.044}$ 
& $0.60^{0.11}_{0.11}$ 
& $13.3^{2.1}_{2.1}$ 
& $<0.66$ 
\\
flat+wk-$h$
& (1.00) 
& $1.08^{0.11}_{0.10}$ 
& $0.018^{0.009}_{0.006}$ 
& $0.15^{0.05}_{0.04}$ 
& $0.44^{0.26}_{0.27}$ 
& $0.56^{0.24}_{0.24}$ 
& $0.059^{0.025}_{0.025}$ 
& $0.58^{0.11}_{0.11}$ 
& $14.1^{1.3}_{1.3}$ 
& $<0.66$ 
\\
flat+wk-$h$+LSS
& (1.00) 
& $1.06^{0.14}_{0.10}$ 
& $0.020^{0.010}_{0.007}$ 
& $0.11^{0.02}_{0.02}$ 
& $0.68^{0.09}_{0.13}$ 
& $0.32^{0.11}_{0.11}$ 
& $0.049^{0.019}_{0.019}$ 
& $0.67^{0.11}_{0.11}$ 
& $14.3^{1.3}_{1.3}$ 
& $<0.63$ 
\\
\\[0.01cm]\tableline\\[0.01cm]
\mCBIofg $(\ell>610)$ & & & & & & & & & & \\
wk-$h$
& $1.06^{0.15}_{0.14}$ 
& $1.14^{0.15}_{0.13}$ 
& $0.068^{0.065}_{0.036}$ 
& $0.15^{0.10}_{0.08}$ 
& $0.44^{0.26}_{0.28}$ 
& $0.71^{0.26}_{0.26}$ 
& $0.264^{0.207}_{0.207}$ 
& $0.59^{0.11}_{0.11}$ 
& $13.1^{1.8}_{1.8}$ 
& $<0.67$ 
\\
flat+wk-$h$
& (1.00) 
& $1.10^{0.12}_{0.10}$ 
& $0.047^{0.056}_{0.019}$ 
& $0.18^{0.07}_{0.07}$ 
& $0.41^{0.22}_{0.26}$ 
& $0.62^{0.23}_{0.23}$ 
& $0.188^{0.166}_{0.166}$ 
& $0.63^{0.11}_{0.11}$ 
& $12.6^{1.4}_{1.4}$ 
& $<0.66$ 
\\
flat+wk-$h$+LSS
& (1.00) 
& $1.05^{0.14}_{0.09}$ 
& $0.041^{0.017}_{0.017}$ 
& $0.13^{0.03}_{0.03}$ 
& $0.66^{0.09}_{0.13}$ 
& $0.35^{0.11}_{0.11}$ 
& $0.082^{0.037}_{0.037}$ 
& $0.71^{0.11}_{0.11}$ 
& $13.1^{1.4}_{1.4}$ 
& $<0.62$ 
\\
\\[0.01cm]\tableline\\[0.01cm]
CBIo200 & & & & & & & & & & \\
wk-$h$
& $1.12^{0.13}_{0.14}$ 
& $1.14^{0.12}_{0.11}$ 
& $0.048^{0.023}_{0.024}$ 
& $0.22^{0.09}_{0.08}$ 
& $0.36^{0.27}_{0.25}$ 
& $0.82^{0.32}_{0.32}$ 
& $0.152^{0.088}_{0.088}$ 
& $0.58^{0.10}_{0.10}$ 
& $12.8^{1.9}_{1.9}$ 
& $<0.67$ 
\\
flat+wk-$h$
& (1.00) 
& $1.07^{0.11}_{0.09}$ 
& $0.025^{0.015}_{0.010}$ 
& $0.19^{0.09}_{0.07}$ 
& $0.41^{0.25}_{0.26}$ 
& $0.61^{0.24}_{0.24}$ 
& $0.071^{0.033}_{0.033}$ 
& $0.63^{0.11}_{0.11}$ 
& $12.6^{1.7}_{1.7}$ 
& $<0.64$ 
\\
flat+wk-$h$+LSS
& (1.00) 
& $1.04^{0.14}_{0.08}$ 
& $0.028^{0.014}_{0.011}$ 
& $0.12^{0.03}_{0.02}$ 
& $0.68^{0.09}_{0.13}$ 
& $0.33^{0.11}_{0.11}$ 
& $0.061^{0.025}_{0.025}$ 
& $0.70^{0.11}_{0.11}$ 
& $13.6^{1.4}_{1.4}$ 
& $<0.59$ 
\\
\\[0.01cm]\tableline\\[0.01cm]
\dCBI & & & & & & & & & & \\
wk-$h$
& $1.09^{0.11}_{0.24}$ 
& $1.16^{0.15}_{0.14}$ 
& $0.078^{0.070}_{0.049}$ 
& $0.21^{0.11}_{0.12}$ 
& $0.42^{0.31}_{0.29}$ 
& $0.85^{0.32}_{0.32}$ 
& $0.261^{0.190}_{0.190}$ 
& $0.61^{0.10}_{0.10}$ 
& $12.2^{1.8}_{1.8}$ 
& $<0.67$ 
\\
flat+wk-$h$
& (1.00) 
& $1.05^{0.12}_{0.11}$ 
& $0.050^{0.072}_{0.034}$ 
& $0.20^{0.09}_{0.14}$ 
& $0.37^{0.26}_{0.25}$ 
& $0.65^{0.24}_{0.24}$ 
& $0.189^{0.188}_{0.188}$ 
& $0.64^{0.11}_{0.11}$ 
& $12.2^{1.9}_{1.9}$ 
& $<0.66$ 
\\
flat+wk-$h$+LSS
& (1.00) 
& $1.03^{0.13}_{0.11}$ 
& $0.055^{0.064}_{0.035}$ 
& $0.13^{0.04}_{0.04}$ 
& $0.55^{0.16}_{0.28}$ 
& $0.50^{0.23}_{0.23}$ 
& $0.187^{0.179}_{0.179}$ 
& $0.66^{0.13}_{0.13}$ 
& $12.9^{2.0}_{2.0}$ 
& $<0.65$ 
\\
\\[0.01cm]\tableline\\[0.01cm]
\DASI+\mCBIofg & & & & & & & & & & \\
wk-$h$
& $1.05^{0.05}_{0.06}$ 
& $1.01^{0.11}_{0.07}$ 
& $0.023^{0.004}_{0.004}$ 
& $0.12^{0.04}_{0.04}$ 
& $0.55^{0.17}_{0.22}$ 
& $0.51^{0.19}_{0.19}$ 
& $0.077^{0.024}_{0.024}$ 
& $0.56^{0.10}_{0.10}$ 
& $15.2^{1.5}_{1.5}$ 
& $<0.63$ 
\\
flat+wk-$h$
& (1.00) 
& $0.99^{0.08}_{0.05}$ 
& $0.021^{0.004}_{0.003}$ 
& $0.14^{0.03}_{0.03}$ 
& $0.56^{0.18}_{0.26}$ 
& $0.46^{0.21}_{0.21}$ 
& $0.057^{0.017}_{0.017}$ 
& $0.62^{0.11}_{0.11}$ 
& $13.9^{0.8}_{0.8}$ 
& $<0.39$ 
\\
flat+wk-$h$+LSS
& (1.00) 
& $1.00^{0.10}_{0.06}$ 
& $0.022^{0.004}_{0.004}$ 
& $0.12^{0.02}_{0.02}$ 
& $0.66^{0.09}_{0.10}$ 
& $0.33^{0.10}_{0.10}$ 
& $0.048^{0.009}_{0.009}$ 
& $0.68^{0.09}_{0.09}$ 
& $13.8^{0.8}_{0.8}$ 
& $<0.42$ 
\\
\\[0.01cm]\tableline\\[0.01cm]
\DASI+Boom+\mCBIofg & & & & & & & & & & \\
wk-$h$
& $1.03^{0.05}_{0.05}$ 
& $0.95^{0.09}_{0.05}$ 
& $0.022^{0.003}_{0.003}$ 
& $0.13^{0.03}_{0.03}$ 
& $0.52^{0.18}_{0.20}$ 
& $0.52^{0.19}_{0.19}$ 
& $0.074^{0.022}_{0.022}$ 
& $0.56^{0.10}_{0.10}$ 
& $15.0^{1.4}_{1.4}$ 
& $<0.52$ 
\\
flat+wk-$h$
& (1.00) 
& $0.94^{0.06}_{0.04}$ 
& $0.021^{0.002}_{0.002}$ 
& $0.14^{0.03}_{0.03}$ 
& $0.55^{0.18}_{0.28}$ 
& $0.48^{0.21}_{0.21}$ 
& $0.058^{0.016}_{0.016}$ 
& $0.61^{0.11}_{0.11}$ 
& $14.0^{0.5}_{0.5}$ 
& $<0.31$ 
\\
flat+wk-$h$+LSS
& (1.00) 
& $0.94^{0.08}_{0.04}$ 
& $0.022^{0.002}_{0.002}$ 
& $0.13^{0.02}_{0.02}$ 
& $0.63^{0.09}_{0.11}$ 
& $0.37^{0.10}_{0.10}$ 
& $0.051^{0.008}_{0.008}$ 
& $0.65^{0.07}_{0.07}$ 
& $13.9^{0.5}_{0.5}$ 
& $<0.36$ 
\\
\\[0.01cm]\tableline\\[0.01cm]
all-data  & & & & & & & & & \\
wk-$h$
& $1.04^{0.05}_{0.05}$ 
& $0.98^{0.10}_{0.06}$ 
& $0.023^{0.003}_{0.003}$ 
& $0.12^{0.03}_{0.03}$ 
& $0.55^{0.17}_{0.20}$ 
& $0.49^{0.18}_{0.18}$ 
& $0.075^{0.023}_{0.023}$ 
& $0.56^{0.11}_{0.11}$ 
& $15.1^{1.4}_{1.4}$ 
& $<0.57$ 
\\
flat+wk-$h$
& (1.00) 
& $0.96^{0.09}_{0.05}$ 
& $0.022^{0.003}_{0.002}$ 
& $0.13^{0.03}_{0.03}$ 
& $0.60^{0.15}_{0.26}$ 
& $0.42^{0.20}_{0.20}$ 
& $0.055^{0.015}_{0.015}$ 
& $0.64^{0.11}_{0.11}$ 
& $13.9^{0.5}_{0.5}$ 
& $<0.35$ 
\\
flat+wk-$h$+LSS
& (1.00) 
& $0.97^{0.09}_{0.05}$ 
& $0.022^{0.003}_{0.002}$ 
& $0.12^{0.02}_{0.02}$ 
& $0.66^{0.09}_{0.12}$ 
& $0.35^{0.10}_{0.10}$ 
& $0.051^{0.008}_{0.008}$ 
& $0.66^{0.08}_{0.08}$ 
& $13.8^{0.5}_{0.5}$ 
& $<0.39$ 
\\
\\
\enddata
\tablecomments{Cosmological parameter estimates as in
Table~\ref{tab:paramscbiO}, except for a variety of data combinations
which test and compare results. Only the wk-$h$, flat+wk-$h$ and
flat+wk-h+LSS priors
are shown. }
\end{deluxetable*}

The effects of discarding the CBI data at $\ell<610$ on the full suite
of parameters and priors can be seen by comparing
Table~\ref{tab:paramscbiO} and Table~\ref{tab:paramscbitests}.  In this
cut, we discarded the first three bins of CBI data. It can be seen
here that the constraints on cosmological parameters degrade
gracefully as data are discarded.  We have also tested the effect of
discarding the first four bins of CBI data (i.e., up to $\ell=750$) and
we find that the uncertainties continue to increase, as expected, but
no large variations in the central values of the parameters are seen.
It is clear, therefore, that the CBI results are robust in this
regard.

\begin{figure}
\plotone{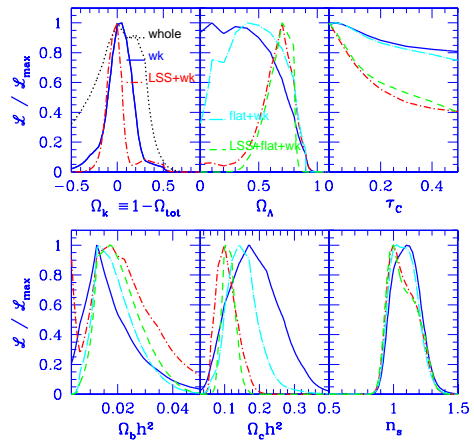}
\caption{Same as Fig.~\ref{fig:likecbimO}, except for
\mCBIefg+\DMR. By comparing this figure with Fig.~\ref{fig:likecbimO},
we see that the particular choice of bin boundaries does not make a significant
difference to the parameter estimation.  This can also be seen from
the
comparison of the even and odd binning cases in
Table~\ref{tab:paramscbitests}. }
\label{fig:likecbimE}
\end{figure}

We have compared the results derived from the two alternate binnings
of the data --- the ``odd'' binning and the ``even'' binning.  The
results are shown in Figure~\ref{fig:likecbimE},
Figure~\ref{fig:12sigbothcbimprs},
and Table~\ref{tab:paramscbitests}.  We see that the derived values
of the parameters agree to within the uncertainties in all cases, and
that the uncertainties are comparable.  This comparison demonstrates
clearly that there is no dependence of the cosmological results on the
binning choice.

\begin{figure}
\plotone{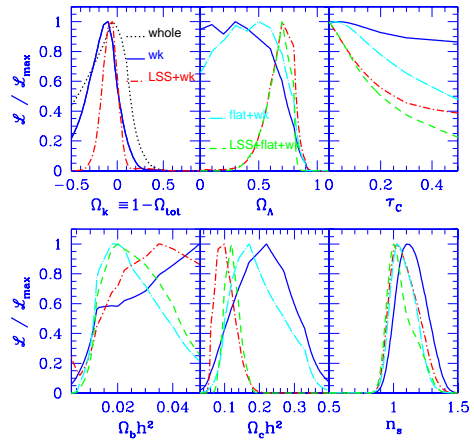}
\caption{Same as Fig.~\ref{fig:likecbimO}, except for the bin size,
which is 200 here. By comparing this figure with
Fig.~\ref{fig:likecbimO}, we see that the bin size does not make a
large difference to the parameter estimation, excepting $\Omega_b
h^2$, for which the $\Delta \ell=200$ binwidth case does not give a
useful measurement.  This can also be seen from the comparison of the
$\Delta \ell =140$ and $\Delta \ell=200$ cases in
Table~\ref{tab:paramscbitests}.}
\label{fig:likecbimO200}
\end{figure}

We have also compared the values of the parameters extracted using bin
widths of $\Delta \ell=140$ (Figure~\ref{fig:likecbimO}) and $\Delta
\ell=200$ (Figure~\ref{fig:likecbimO200}; see also
Table~\ref{tab:paramscbiO} and Table~\ref{tab:paramscbitests}).
Comparison of these figures and the actual values and associated
uncertainties shows that the cosmological results are in excellent
agreement.  The results are not strongly dependent on the bin width,
although there is, of course, some loss of information at the larger
bin width, which is reflected in the larger uncertainties. 

We have also tested the effect of varying the residual source
contribution on parameters. As expected from its small relative
contribution, we find very little difference if we either assume our standard
power value described in Papers II and III, allow for a 50\% error in
that estimate, or multiply the standard power by a factor of 2.25. 

We end our discussion of tests of the cosmological parameter
extraction from the CBI observations by  showing projections of
the full seven-dimensional likelihood function onto  2D contour plots
of the likelihoods of various combinations of cosmological parameters
$\{ \omega_{\rm cdm}, \Omega_k, \omega_b,\Omega_\Lambda, n_s\}$ to
 illustrate further  the consistency of the various
\cbi\ data subsets used throughout this work.  These 2D
contour plots provide a different insight into the tests of data
consistency than do the 1D plots of the previous subsections, which
can be illuminating.  For example, they can show directly whether
there are isolated multiple peaks in the likelihood surface, whereas
this information is lost in the marginalization of the 1D plots. 

 In Paper III we showed the power spectra for the
individual mosaic fields, and we found that the differences are not
statistically significant.  
Here we derive constraints on the cosmological parameters from
three sets of pairwise-combined mosaic fields. These give stronger
constraints on parameters than the single fields and thus provide a
stronger consistency check. 
We find there is good consistency between the three
pairs, showing that no field is seriously discrepant with the other
two, in agreement with our finding in Paper III that there are no
significant differences between the spectra from the three fields.
The agreement between the three pairs of fields can be seen in
Figure~\ref{fig:12sigbothcbimprs}, the results of the pairwise
splitting of the three mosaic data sets.  Displayed here are the 1-
and 2-$\sigma$ contours for the odd binning of the $\Delta \ell =140$
data.

\begin{figure*}
\epsscale{0.8}
\plotone{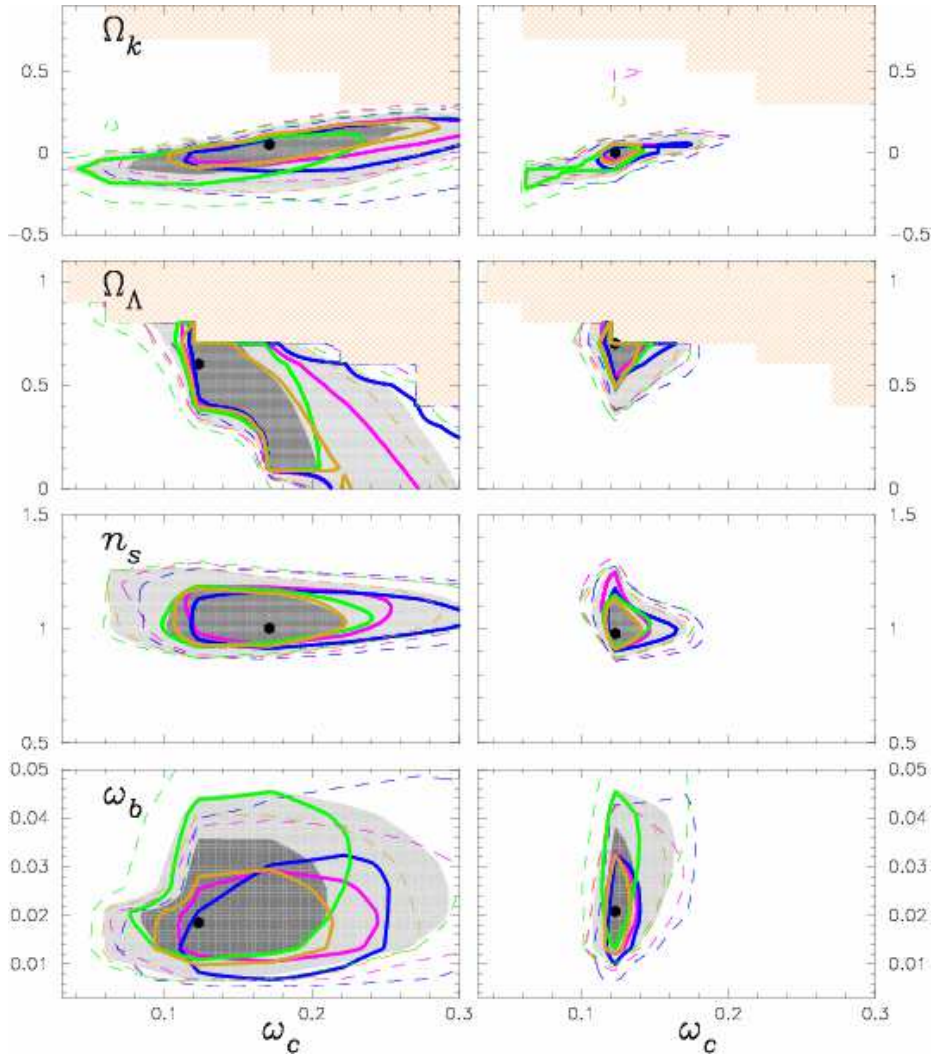}
\caption{Comparison of likelihood contours for the three pairwise
combinations of the three mosaic fields for the \mCBIofg + \DMR data.
Full lines are $1\sigma$ and dashed lines are $2\sigma$ contours.  The
hatched regions indicate portions excluded by the range of parameters
considered (see Table~\ref{tab:grid}).  The weak-$h$ prior applies in
the $\omega_{\rm cdm}$--$\Omega_k$ panel and flat+weak-$h$ prior
applies in the other panels.  The plots on the left do not have the
LSS prior, while those on the right do.  The pairs of mosaics are as
follows: 02h+20h (magenta); 02h+14h (dark blue); 14h+20h (green).  The
results for the three fields combined are shown by the dark grey
regions ($1\sigma$) and the light grey regions ($2\sigma$).  For
comparison, the results for the alternate binning (\mCBIefg + \DMR)
are shown by the gold contours.  Comparison of the plots on the left
with those on the right show the dramatic impact of the LSS prior on
the $\omega_{\rm cdm}$ determination, its role in $\Omega_\Lambda$
determination, and its relatively small impact in the other variable
directions.  The dramatic reduction in the uncertainties in
$\omega_{\rm cdm}$ is due to the combined effects of the CBI data and
the LSS prior (see text). (For this comparison, the slightly smaller
database in \citep{lange01} was used rather than that in
Table~\ref{tab:grid}.) }
\label{fig:12sigbothcbimprs}
\end{figure*}

 The effect of introducing the LSS prior is also shown in
Figure~\ref{fig:12sigbothcbimprs}. It primarily reduces the
uncertainties in $\omega_{\rm cdm}$.  Note that it is the {\it combination} of
the CBI observations and the LSS prior that accomplishes this.  This can be
seen by comparing Figure~\ref{fig:likedmr}, in which the uncertainties
under the LSS prior are large, to Figure~\ref{fig:likecbimO} in which
the CBI data have been included and the uncertainties are greatly reduced.

A 2D comparison of the parameters extracted under the two binning
schemes is shown in Figure~\ref{fig:12sigbothcbimprs}, with the
effect of adding the LSS prior also shown.  These
should be compared with Figure~\ref{fig:likecbimO} and
Figure~\ref{fig:likecbimE}, and with the values given in
Table~\ref{tab:paramscbitests}.  We see here, once again, that the
consistency between these two binning schemes is excellent.

One of the caveats that is important to bear in mind when dealing with
7D-plus parameter spaces is that the limits derived from projection
onto one or two directions in parameter space by full Bayesian
marginalization (integration) over the other variables can be
misleading in certain cases. For example, highly likely
models which exist far from broad likelihood peaks may be ruled
out.  
With all of the CMB data, many of the variables are well
localized and this is not a big problem. However, near degeneracies
among cosmological parameters do exist in these inflation model spaces
\citep{eb98}. The correlations can be quantified by considering
parameter eigenmodes \citep[e.g.,][]{bh95,bet,eb98,lange01} which yield
linear combinations of the ${\cal C}_\ell$-database variables that
give orthogonal errors in the neighborhood of the maximum likelihood
values.

For \mCBIofg+\DMR\ with weak-$h$ prior, two combinations are
determined to better than 10\% and two others to 15\%. The three
best-determined values involve a predominantly $n_s$ combination, with
$\tau_C$ and amplitude mixed in, and then, somewhat remarkably, a
predominantly $\Omega_k$ combination, followed by a mix of many
parameters, including amplitude, $\Omega_\Lambda$ and
$\omega_{cdm}$. With all-data, weak-$h$ gives 4 combinations to better
than 10\%. The first two are different mixes of $n_s$ and $\Omega_k$,
next is mainly $\omega_b$ and then a $\omega_{\rm
cdm}$--$\omega_{b}$--$\Omega_\Lambda$ combination. LSS sharpens up the
fourth eigenmode especially.

In conclusion we would emphasize the excellent overall consistency of
the cosmological parameters extracted from the CBI mosaic and deep
observations.  It was shown in Papers II and III that various subsets of
the data are consistent with each other to within the levels expected,
given the uncertainties.  Here we have shown that the cosmological
parameters derived from different subsets of the data, and from
different binning schemes and bin widths, are also self-consistent.
We conclude that both the CBI data themselves and the cosmological
parameter extraction from the CBI observations are robust. 

\section{Optimal Power Spectra} \label{sec:optimal1}

In this section, we apply the parameter-estimation methods of
Appendix \ref{sec:ps} to combine power spectra derived with different
bandpowers and with different window functions onto a common set of
bands. A byproduct is the determination of how consistent the power
spectra are and the values of various experimental parameters that can
be adjusted to increase agreement  (e.g., adjustments to the
flux density scale). For all of the bandpower applications, a
best-fit model was used for ${\cal C}_\ell^{(s)}$ rather than flat
bandpowers.

\subsection{CBI Optimal Power Spectra}

 In addition to the extraction of the optimal power spectrum for the
whole CBI data set, we are interested in extracting optimal power
spectra for subsets of the data.  Although these optimal spectra are
not used in the estimation of the cosmological parameters, they do
provide an invaluable means of comparing the various data sets, and,
of course, of comparing the CBI data to other CMB data sets.

We begin the extraction of optimal spectra by examining the pairwise
combinations of the three mosaic fields.  Thus we combine pairs of the
three mosaic fields (02\hour, 14\hour, and 20\hour).  These are
 separated by about 6\hour\
in RA,  which  is sufficient to treat them as
independent fields and uncorrelated data sets. When combining the
fields we do not include separate calibration errors since
 these are  common to all fields, and we assume no error
on the estimate of the residual source component. The NVSS
sources are  projected out as described in Papers II and
III, and that is included in this treatment.

\begin{figure*}
\epsscale{0.8}
\plotone{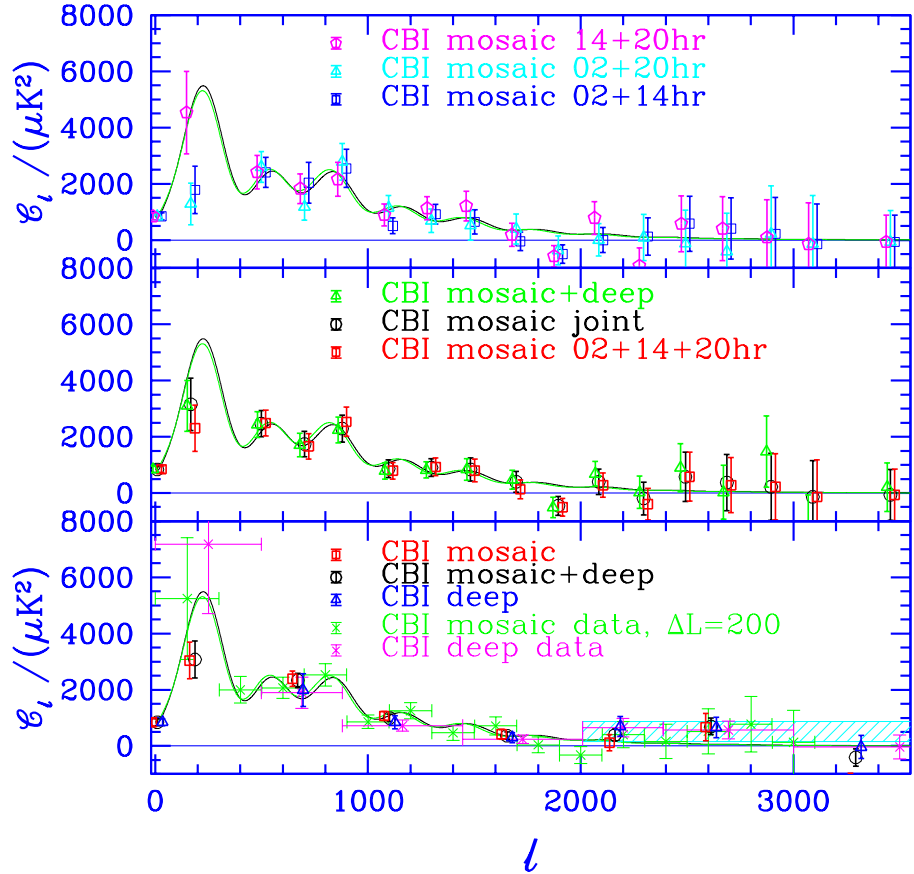}
\label{fig:compare3mos}
\figcaption{Optimal power spectra from CBI observations.
{\it Upper panel:}
Consistency between pairs of \mCBIofg\ fields. Here the \mCBIofg\ data
have been re-binned to the \mCBIocg\ bands for the purpose of
comparison with other data sets (see text). At low-$\ell$
the errors are dominated by sample variance, which is a fraction of
the fitted band-power --- thus the apparent discrepancy between the pairs
in the lowest-$\ell$ bin, due to the 02h field, is not significant (see
Paper III).  
The best fit model, under the weak-$h$ and  flat+weak-$h$ priors, 
to the \DMR+\mCBIofg\ data (black curve) has
$\{\Omega_{\rm {tot}}$, $\Omega_\Lambda$, $\Omega_b h^2$, $\Omega_{\rm cdm}
h^2$, $n_s$, $\tau_C\}$=$\{1.0, 0.6, 0.0225, 0.12, 0.95, 0.025\}$. 
The best fit model, under the weak-$h$,  flat+weak-$h$ and
flat+weak-$h$+LSS priors, for ALL-data (magenta curve) has
parameters $\{1.0, 0.5, 0.02, 0.14, 0.925, 0\}$. 
{\it Middle panel:}
Combined
joint 3-field \mCBIofg\ and \dCBI\ power spectra denoted by the green
triangles; black circles denote the joint 3-field \mCBIofg\ spectrum
and red squares denote the 3 \mCBIofg\ fields when combined as with the pair
cases.  The results are stable independent of whether even or odd
binning is used.
{\it  Lower panel:}
The \mCBIocg\ and \dCBI\ data are shown
together with optimal spectra for deep, mosaic and deep+mosaic on the
standard \dCBI\ bands.}
\end{figure*}

The upper panel of Figure~\ref{fig:compare3mos} compares the combined
spectra for the three pairs of mosaic fields. The center
panel shows the combination of the three individual field spectra, the
spectrum obtained from the joint analysis of the three fields
(Paper~III), and the combination of all six mosaic and deep
fields.  We
have separated the upper and center panels for clarity, but when
plotted together they show excellent consistency between all the
combinations we have considered.   In the spectra shown here, the
\mCBIofg\ data have been used and combined onto the $\Delta \ell$=200
bins.   We find excellent agreement between the extracted
optimal spectra, regardless of whether the \mCBIefg\ or \mCBIocg\ data
are used. This shows not only
that the extracted data are self-consistent, but also that the
band-to-band correlations are being treated correctly. We see from the
upper panel of Figure~\ref{fig:compare3mos}   that the agreement of
the pairs is robust. This result is related to that obtained by
comparing the power spectra for each individual mosaic field shown in
Paper III. These show excursions among the power spectra for
individual fields,  although, as shown in Paper III,
these excursions are not statistically significant. The
pair-combined spectra shown here, with their increased statistical
weights, indicate again that the excursions are compatible with
expectations.

The three deep fields are single pointings with long integration
times.  Since the deep observations have better signal-to-noise ratios
in the $\ell > 2000$ range, it is useful to combine the deep spectra
with the mosaic spectra, which have less cosmic variance at low
$\ell$. One point to note is that two of the deep fields are embedded
in the mosaic fields.  However, the data used for the mosaics are only
a subset of the data used in the the deep fields analysis,
corresponding to the typical integration time on an individual mosaic
pointing, which is about four hours.  We therefore expect the
correlations between the deep and mosaic data to be small and
we ignore them when forming the optimal spectra. The center panel of
Figure~\ref{fig:compare3mos} compares the combined \dCBI\ and joint
\mCBIofg\ data with those from just the joint \mCBIofg. Both cases use
top hat window functions and are mapped onto the $\Delta \ell=200$
mosaic binning. The excess power anomaly seen in the deep data is not
that evident with these relatively narrow bands, and the basic result
is good agreement between the two.

The lower panel of Figure~\ref{fig:compare3mos} compares the mosaic,
deep, and mosaic+deep optimal power using the much coarser \dCBI\
$\Delta \ell$ binning.  This is similar to Figure~14 of
Paper~III, for which the mosaic data were evaluated directly in the
\cbi-pipeline analysis onto the \dCBI\ bands.  Here we have obtained a
similar result by using only the \mCBIofg\ data: although there is no
detection of mosaic power at $\ell \sim 2200$, there is agreement in
enhanced power levels at $\ell \sim 2700$, although the larger
$1\sigma$ mosaic confidence region is consistent with a
``non-detection''. The joint deep+mosaic bandpower has slightly
smaller errors than the deep-only case. 

The combined spectrum shows a clear detection of the expected damping
of the power out to $ \ell \sim 2000$. Thus the unique experimental
setup of the CBI has further validated the cosmological paradigm outlined in
\S~\ref{sec:intro} by confirming one of the key ingredients.
Note the agreement in the excess power in the $\ell \sim 2700$ band
(though errors differ).  Further study is needed to confirm this
excess, and, if confirmed, to determine the
source of the power (see Paper VI for further discussion).
 
\subsection{The Overall Optimal Spectrum}\label{sec:optimal2}

In this section we compare the ${\cal C}_b$ (band-power) spectrum from the CBI with
the spectra obtained from some other CMB experiments, and we then
combine the CBI data with the data from the \Boomerang, \DASI, \Maxima,
and \VSA\ experiments in order to obtain a new optimal ${\cal C}_b$
spectrum out to $\ell=3500$.  This represents a considerable extension
of the optimal spectrum beyond the previous limit at $\ell \sim 1000$.

\begin{figure*}
\epsscale{0.8}
\plotone{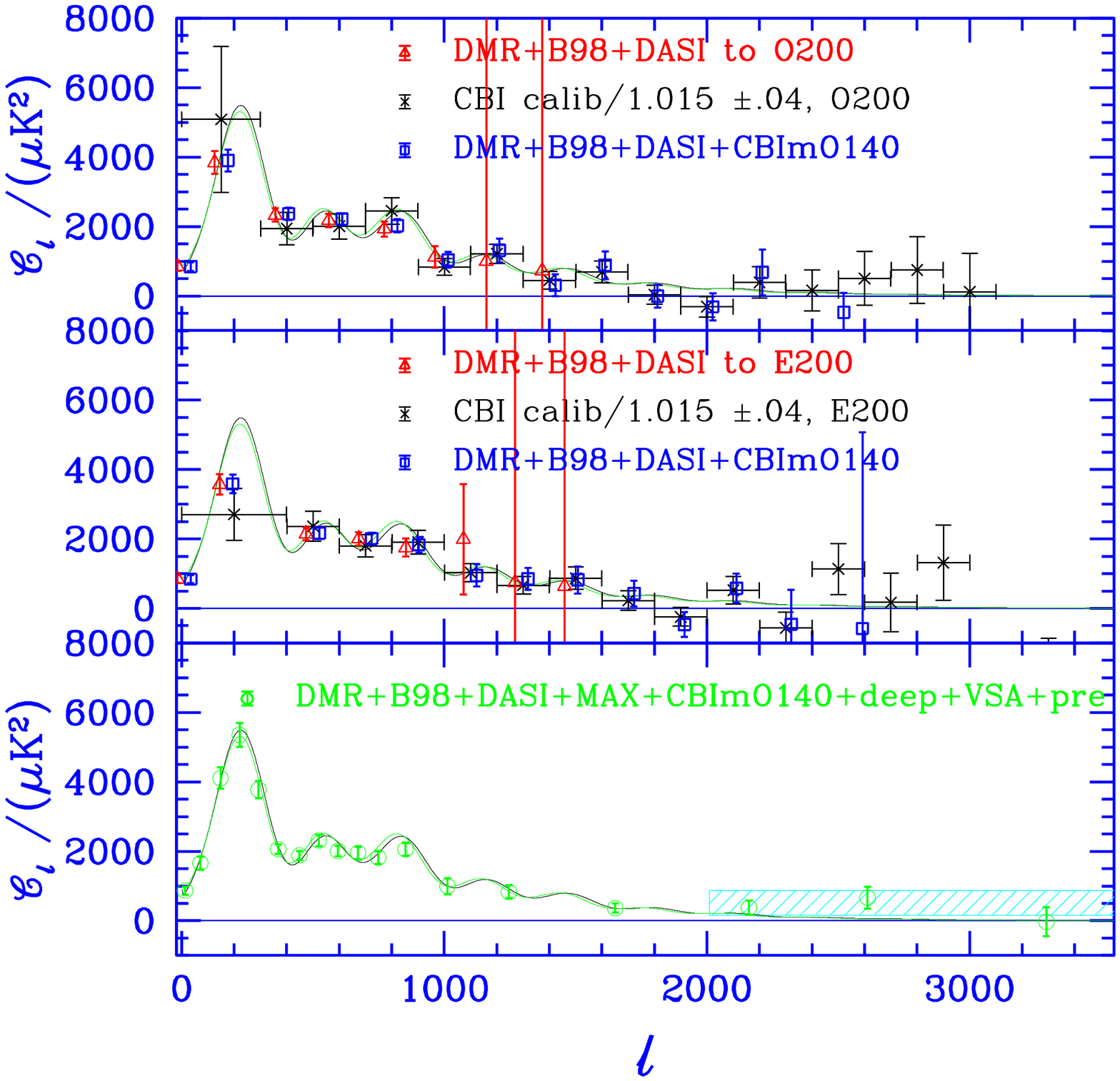}
\caption{Optimal power spectra: \cbi\ compared to \Boomerang\ and
\DASI.  
{\it Upper panel:}
Joint \mCBIocg\ power spectrum
compared with optimal power spectra using \Boomerang+\DASI\ data (red
triangles) and \mCBIofg+\Boomerang+\DASI\ data (dark blue circles) in the
\mCBIocg\ binning. The optimal spectra correspond to a
maximum-likelihood fit to the power in bands for the three spectra,
marginalized over beam and calibration uncertainty.  The best fit
curves shown are the same as in Fig.~\ref{fig:compare3mos}.
{\it  Middle panel:}
same as the upper panel, but using the \mCBIecg\ binning.
{\it Lower panel:}
The optimal spectrum for ``all-data'', with a finer
$\Delta \ell =75$ binning up to $\sim 800$, and deep binning at large
$\ell$. The $\ell > 2000$ excess found with the \dCBI\ data is denoted
by the light blue hatched region (95\% confidence limit).
}
\label{fig:comparecbidasib98}
\end{figure*}

A comparison of the pre-CBI optimal spectrum with the optimal spectrum
including the CBI deep and mosaic data is shown in
Figure~\ref{fig:comparecbidasib98}.  As noted above, each experiment has
unique binnings, band-to-band correlations, and calibration and beam
errors.  This makes a straightforward visual comparison difficult. To
facilitate such comparisons, we therefore construct the optimal power
spectrum for these experiments binned onto the $\Delta \ell =200$
\cbi\ bins. Sample results are shown in
Figure~\ref{fig:comparecbidasib98}, along with some best-fit models.  We
see here that the agreement between the different CMB experiments is
excellent. In the derivation of the optimal spectra shown here
calibration uncertainties of 10\% for \Boomerang, 4\% for \DASI, 3.5\% for \VSA, and
4\% for \Maxima\ were included, together with beam uncertainties of
14\% for \Boomerang\  and 5\% for \Maxima.  For this purpose we need to
include the calibration uncertainty for \cbi, which we take to be a
conservative 5\%. These are all incorporated in the spectra, which
leads to significant correlations among bands associated with the beam
uncertainty. That is, one must be careful in using optimal spectra
directly for parameter estimation since, although we can compute the
Fisher matrix, we do not know the likelihood surface in detail.

We find the optimal power spectrum for \Boomerang, \DASI, and \cbi\
requires a decrease in the temperature calibration by a factor
$1.015\pm 0.04$ for \cbi\ and $1.05\pm 0.05$ for \Boomerang, and an
increase by a factor $1.01\pm 0.04$ for \DASI. The data also favor
an increase in the \Boomerang\ beam by $1.07\pm 0.04$ for the odd binning
and $1.05\pm 0.04$ for the even binning. These values are consistent
with the quoted errors in all cases. Indeed, assuming that the power
spectra are derived from a single underlying spectrum, which the
different experiments are sampling in their respective regions of the
sky, one could use this technique to determine the calibrations and
beams. For example, we note that the determinations for \Boomerang\
have errors significantly smaller than the quoted uncertainties once
they have been compared with the two sets of interferometer data,
which have no beam uncertainties and much smaller calibration
uncertainties.

An important caveat to these optimally-combined ${\cal C}_b$'s is that
power spectra for the individual experiments were assumed to be
independent. We have already commented on this for \cbi\ deep and
mosaic combinations. In addition \DASI's fields overlap about 5\% of
the \Boomerang\ area, so there is correlation between \Boomerang\ and
\DASI\ which is not taken into account in this treatment. The
correlation would have to be addressed in order to claim absolute
accuracy in the adjustment of experimental parameters.

Figure~\ref{fig:comparecbidasib98} shows very good agreement in the
$300\lesssim \ell \lesssim 1000$ overlap region for \Boomerang\ and
\DASI\ combined onto \cbi\ points, and with all 3 experiments
combined. (Note that the window functions for the combined spectra are
top hats and not the $W_{B\ell}$ of the data points shown.) We also
find the combined spectrum looks quite similar when all-data are used.

The method we used for Figure~\ref{fig:comparecbidasib98} has also been
applied to construct optimal bandpowers for all of the above data
(including \mCBIofg\ and \dCBI) with  finer binning  for
$\ell < 1000$.  This variable binning makes use of the high quality
\Boomerang\ data with its intrinsic $\Delta \ell =50$ spacing
 at low $\ell$ and it also makes use of the high-$\ell$
coverage of the CBI (out to $\ell=3500$). The spectrum is compared
in Paper VI with power spectra computed for the Sunyaev-Zeldovich
effect, and will not be discussed further here.

\section{Cosmological Parameter Estimates from Combined CMB Data}\label{sec:allparams}

We test the consistency of our parameter estimations by comparing with
different combinations of \cbi\ and other data sets. We also derive
estimates for the parameters from the full compilation of data
available.

\begin{figure*}
\epsscale{0.8}
\plotone{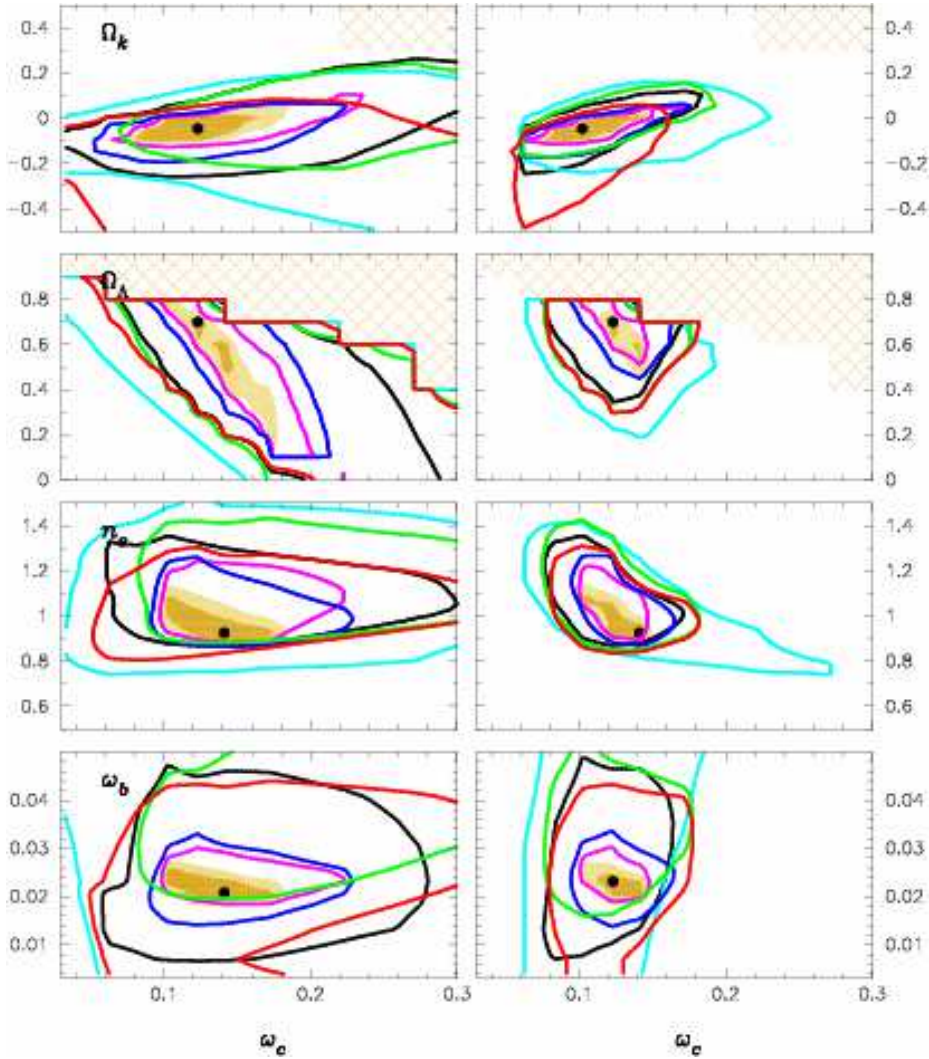}
\figcaption{Comparison of different experiments.  2-$\sigma$ likelihood
contours for the weak-$h$ prior ($\omega_{\rm cdm}$--$\Omega_k$ panel) and
flat+weak-$h$ prior for the rest, for the following CMB experiments in
combination with \DMR: \mCBIefg\ (black), \Boomerang\ (magenta),
\DASI\ (dark blue), \Maxima\ (green), \VSA\ (red) and ``prior-CMB" =
\Boomerang-NA+\TOCO+Apr99 data (light blue). Light brown region shows the
2-$\sigma$ contour when all of the data are taken together, dark brown shows
the 1-$\sigma$ contour. The LSS prior has not been used in deriving
the plots on the left, but it has for those on the right. The hatched
regions indicate portions excluded by the range of parameters
considered (see Table~\ref{tab:grid}). This figure shows great
consistency as well as providing a current snapshot of the collective
CMB data results. Even without the LSS prior (or the HST-$h$ or SN1a
priors), localization of the dark matter density is already occurring,
but $\Omega_\Lambda$ still has multiple solutions. The inclusion of
the SN1a and/or the HST-$h$ priors does not concentrate the bull's-eye
determinations much more for the all-data shaded case. Note that the
expectation of minimal inflation models is that $\Omega_k \approx 0$,
$n_s \approx 1$ (usually a little less). The Big Bang Nucleosynthesis
result, $\omega_b = 0.019 \pm 0.002$ also rests comfortably within the
bull's-eye. 
As expected from
the results in Table~\ref{tab:paramscbiOall}, relaxing the flat
criterion has little impact.\label{fig:2sigbothall}}
\end{figure*}

The consistency in cosmological parameter space at the $2\sigma$ level
among these 21 experiments, and the 5 higher precision ones, is
illustrated in Figure~\ref{fig:2sigbothall}.
This shows that using \mCBIofg\ data with \DMR\ is quite
comparable at the $2\sigma$ level with what was achieved by
\Boomerang\ (with the smallest errors), \DASI, \Maxima, and \VSA. The
combined data, with the LSS prior applied as well, gives the bull's-eye
determination of Figure~\ref{fig:2sigbothall}.

Table~\ref{tab:paramscbitests} also shows the sequence obtained when
we add data from \cbi's sister interferometry experiment, \DASI, to
\mCBIofg+\DMR. We then add the \Boomerang\ data in order to check for
any effects arising from the combination of data from completely
different experimental setups.  Finally we show the all-data
combination. 

\begin{figure}
\epsscale{0.9}
\plotone{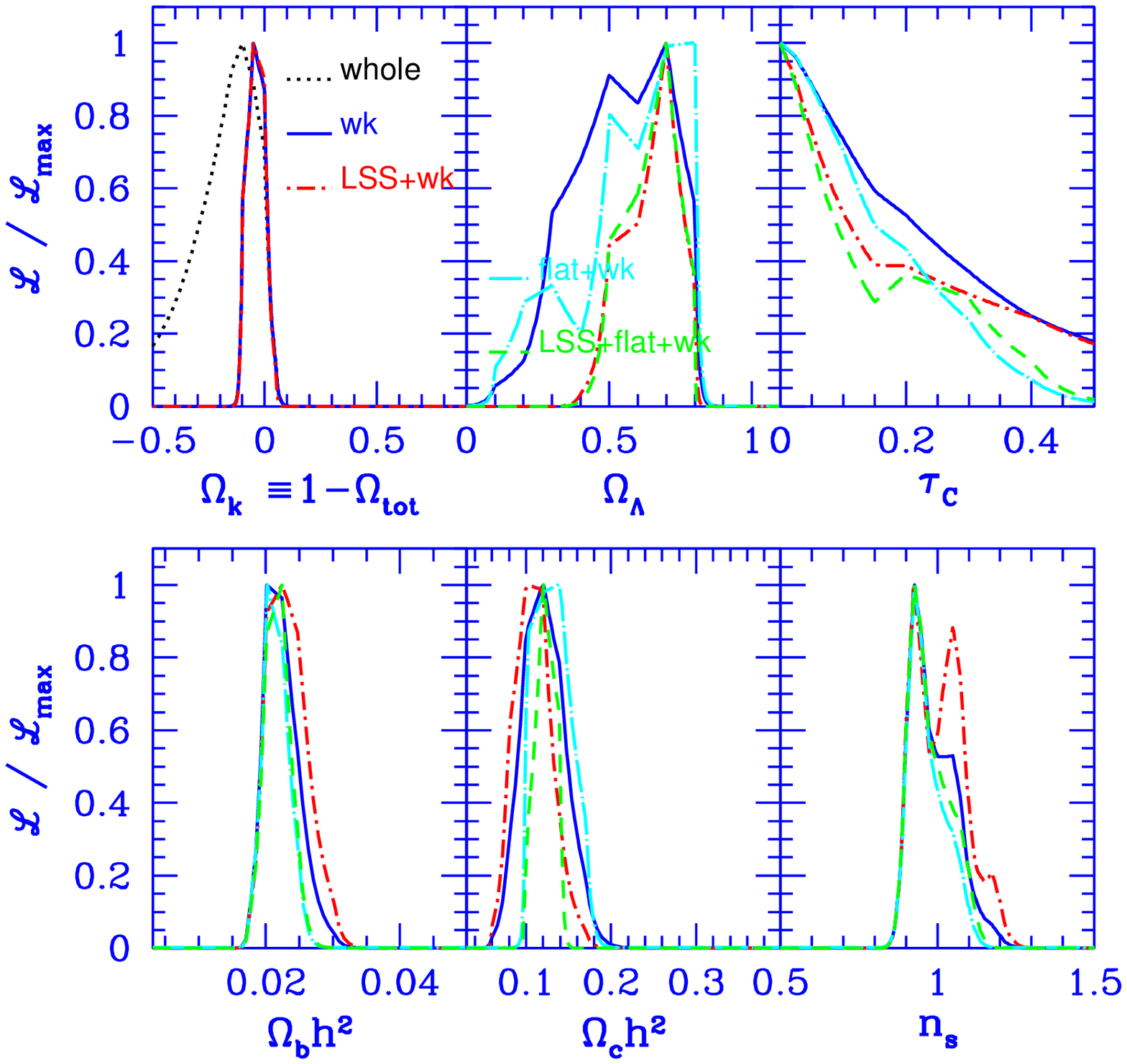}
\caption{1D projected likelihood functions as in Fig.~\ref{fig:likecbimO},
except calculated using all-data (see Table~\ref{tab:paramscbitests}). }
\label{fig:likecbiall}
\end{figure}

\begin{figure}
\plotone{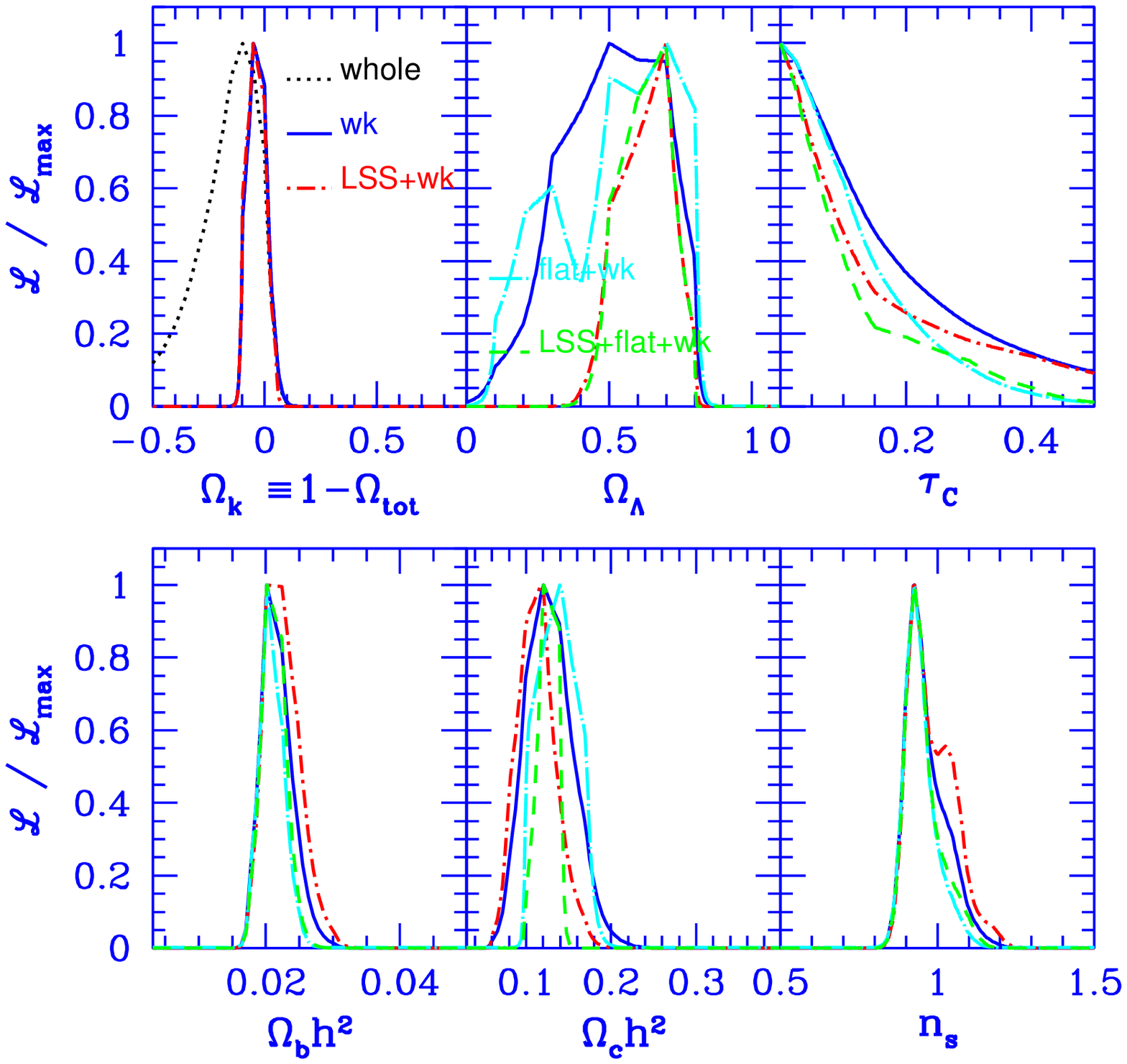}
\caption{ 1D
projected likelihood functions as in Fig.~\ref{fig:likecbimO}, except
calculated using \mCBIofg+\DMR+\DASI+\Boomerang+\VSA\
(the data for \Boomerang\ are taken from \citealt{ruhl02}). }
\label{fig:likecbidasiB98}
\end{figure}

\begin{deluxetable*}{lllllllllll}
\tabletypesize{\scriptsize}
\tablecaption{all-data
\label{tab:paramscbiOall}}
\tablewidth{0pt}
\tablehead{\colhead{Priors}
& \colhead{$\Omega_{\rm tot}$}
& \colhead{$n_s$}
& \colhead{$\Omega_bh^2$}
& \colhead{$\Omega_{\rm cdm}h^2$}
& \colhead{$\Omega_{\Lambda}$}
& \colhead{$\Omega_m$}
& \colhead{$\Omega_b$}
& \colhead{$h$}
& \colhead{Age}
& \colhead{$\tau_c$}
}
\startdata
wk-$h$
& $1.04^{0.05}_{0.05}$ 
& $0.98^{0.10}_{0.06}$ 
& $0.023^{0.003}_{0.003}$ 
& $0.12^{0.03}_{0.03}$ 
& $0.55^{0.17}_{0.20}$ 
& $0.49^{0.18}_{0.18}$ 
& $0.075^{0.023}_{0.023}$ 
& $0.56^{0.11}_{0.11}$ 
& $15.1^{1.4}_{1.4}$ 
& $<0.57$ 
\\
wk-$h$+LSS
& $1.04^{0.05}_{0.04}$ 
& $1.01^{0.09}_{0.09}$ 
& $0.023^{0.004}_{0.003}$ 
& $0.11^{0.03}_{0.03}$ 
& $0.66^{0.09}_{0.13}$ 
& $0.39^{0.11}_{0.11}$ 
& $0.069^{0.020}_{0.020}$ 
& $0.60^{0.09}_{0.09}$ 
& $15.2^{1.5}_{1.5}$ 
& $<0.60$ 
\\
wk-$h$+SN
& $1.03^{0.05}_{0.04}$ 
& $1.03^{0.08}_{0.08}$ 
& $0.024^{0.003}_{0.003}$ 
& $0.11^{0.02}_{0.02}$ 
& $0.71^{0.06}_{0.07}$ 
& $0.32^{0.08}_{0.08}$ 
& $0.061^{0.020}_{0.020}$ 
& $0.64^{0.09}_{0.09}$ 
& $14.8^{1.6}_{1.6}$ 
& $<0.63$ 
\\
wk-$h$+LSS+SN
& $1.03^{0.05}_{0.04}$ 
& $1.04^{0.08}_{0.08}$ 
& $0.024^{0.004}_{0.003}$ 
& $0.10^{0.02}_{0.03}$ 
& $0.71^{0.06}_{0.06}$ 
& $0.33^{0.06}_{0.06}$ 
& $0.064^{0.020}_{0.020}$ 
& $0.63^{0.09}_{0.09}$ 
& $15.0^{1.6}_{1.6}$ 
& $<0.63$ 
\\
\\[0.01cm]\tableline\\[0.01cm]
flat+wk-$h$
& (1.00) 
& $0.96^{0.09}_{0.05}$ 
& $0.022^{0.003}_{0.002}$ 
& $0.13^{0.03}_{0.03}$ 
& $0.60^{0.15}_{0.26}$ 
& $0.42^{0.20}_{0.20}$ 
& $0.055^{0.015}_{0.015}$ 
& $0.64^{0.11}_{0.11}$ 
& $13.9^{0.5}_{0.5}$ 
& $<0.35$ 
\\
flat+wk-$h$+LSS
& (1.00) 
& $0.97^{0.09}_{0.05}$ 
& $0.022^{0.003}_{0.002}$ 
& $0.12^{0.02}_{0.02}$ 
& $0.66^{0.09}_{0.12}$ 
& $0.35^{0.10}_{0.10}$ 
& $0.051^{0.008}_{0.008}$ 
& $0.66^{0.08}_{0.08}$ 
& $13.8^{0.5}_{0.5}$ 
& $<0.39$ 
\\
flat+wk-$h$+SN
& (1.00) 
& $0.99^{0.07}_{0.06}$ 
& $0.023^{0.002}_{0.002}$ 
& $0.12^{0.02}_{0.02}$ 
& $0.71^{0.06}_{0.07}$ 
& $0.29^{0.07}_{0.07}$ 
& $0.045^{0.006}_{0.006}$ 
& $0.71^{0.06}_{0.06}$ 
& $13.6^{0.3}_{0.3}$ 
& $<0.37$ 
\\
flat+wk-$h$+LSS+SN
& (1.00) 
& $0.99^{0.07}_{0.06}$ 
& $0.023^{0.002}_{0.002}$ 
& $0.12^{0.01}_{0.01}$ 
& $0.70^{0.05}_{0.06}$ 
& $0.30^{0.06}_{0.06}$ 
& $0.047^{0.005}_{0.005}$ 
& $0.69^{0.05}_{0.05}$ 
& $13.7^{0.3}_{0.3}$ 
& $<0.39$ 
\\
\\[0.01cm]\tableline\\[0.01cm]
flat+HST-$h$
& (1.00) 
& $0.99^{0.07}_{0.06}$ 
& $0.023^{0.002}_{0.002}$ 
& $0.12^{0.02}_{0.02}$ 
& $0.70^{0.08}_{0.11}$ 
& $0.30^{0.10}_{0.10}$ 
& $0.047^{0.009}_{0.009}$ 
& $0.70^{0.08}_{0.08}$ 
& $13.6^{0.4}_{0.4}$ 
& $<0.37$ 
\\
flat+HST-$h$+LSS
& (1.00) 
& $0.99^{0.08}_{0.06}$ 
& $0.023^{0.002}_{0.002}$ 
& $0.12^{0.02}_{0.02}$ 
& $0.69^{0.06}_{0.08}$ 
& $0.31^{0.08}_{0.08}$ 
& $0.048^{0.006}_{0.006}$ 
& $0.69^{0.06}_{0.06}$ 
& $13.7^{0.3}_{0.3}$ 
& $<0.39$ 
\\
flat+HST-$h$+SN
& (1.00) 
& $0.99^{0.07}_{0.05}$ 
& $0.023^{0.002}_{0.002}$ 
& $0.12^{0.01}_{0.02}$ 
& $0.71^{0.06}_{0.05}$ 
& $0.28^{0.06}_{0.06}$ 
& $0.045^{0.006}_{0.006}$ 
& $0.71^{0.05}_{0.05}$ 
& $13.6^{0.2}_{0.2}$ 
& $<0.37$ 
\\
flat+HST-$h$+LSS+SN
& (1.00) 
& $1.00^{0.06}_{0.05}$ 
& $0.023^{0.002}_{0.002}$ 
& $0.12^{0.01}_{0.01}$ 
& $0.70^{0.05}_{0.05}$ 
& $0.30^{0.05}_{0.05}$ 
& $0.047^{0.004}_{0.004}$ 
& $0.69^{0.04}_{0.04}$ 
& $13.7^{0.2}_{0.2}$ 
& $<0.38$ 
\\
\enddata
\tablecomments{Cosmological parameter estimates as in Table~\ref{tab:paramscbiO},
but now for all-data. }
\end{deluxetable*}

A full suite of priors for this all-data case is given in
Table~\ref{tab:paramscbiOall} and the corresponding 1D likelihood
plots are shown in
Figure~\ref{fig:likecbiall}. Figure~\ref{fig:likecbidasiB98} shows the
small difference when we use \Boomerang+\DASI+\mCBIofg+\DMR. These
combined data yield parameters consistent with those derived
individually: the curvature is close to flat, the spectral index is
close to unity, and the baryon density is near that favoured by Big
Bang nucleosynthesis, $\omega_b =0.019\pm0.002$~\citep{olive,burles,tytler}.

The tables also show estimates for what we term ``derived''
parameters. These are parameters that can be expressed as functions of
our ${\cal C}_\ell$-database parameters. The combinations are
$\Omega_m$, $\Omega_b$, $h$, and the cosmological age $t_0$. We
calculate the means and variances of these functions over the full
probability distribution. We have also applied the same method to the
computation of the statistics of $\ell_{{\rm pk},j}$, ${\cal C}_{{\rm
pk},j}$, and $\ell_{{\rm dip}, j}$, ${\cal C}_{{\rm dip},j}$, and the
position and amplitude of the $j$th peak and $j$th dip in ${\cal
C}_\ell$. 
These are discussed in detail in the following section
(\S~\ref{sec:damp}) and the results from the application to the mosaic
data are shown in Paper III.  
We also determine the values of $\Omega_m h$, $\Gamma$,and
$\Gamma_{\rm eff}$ in a similar fashion, as reported in Paper
VI. Paper VI also reports on calculations of alternate amplitude
parameters to ${\cal C}_{10}$, in particular for the $\sigma_8$
amplitude used in the LSS prior.

The results set out in the tables also show how applying the prior
restrictions LSS, SNIa, or HST-$h$ to the CMB data gives compatible
results.

\newpage
\section{Reheating, Recombination, and Damping} \label{sec:damp}

In this section, we summarize the basic physical effects expected to
have an impact on the ${\cal C}_\ell$ spectra we observe, working back
from now, through the reionization of the universe,
\S~\ref{sec:reheat}, and to the critically important regime for CBI
observations, recombination. We discuss briefly the history of
computations of the anisotropy in \S~\ref{sec:recombepoch}, but our
main goal is to use the parameters of our best-fit cosmological models 
to evaluate the physical and multipole scales characterizing the
decoupling epoch in \S~\ref{sec:recombdamp}. 

\subsection{Reheating} \label{sec:reheat}

We have used the depth to Thomson scattering from a time $t_1$ to the
present $t_0$, $\tau_C = \int_{t_1}^{t_0} c\,dt \,\bar{n}_e \sigma_T$, as
one of our major cosmological parameters, where $\bar{n}_e(t)$ is the
average electron density, $\sigma_T$ is the Thomson cross section and
$c$ is the speed of light. If we assume the Universe has been fully
ionized below a redshift $z_{\rm reh}$, we have
\begin{equation} \tau_C \sim 0.1 (\omega_b/0.02)
(\omega_m/0.15)^{-1/2} ((1+z_{\rm reh})/15)^{3/2}.
\end{equation}
A minimum value for $z_{\rm reh}$ is $\sim 5$, so $\tau_C \gta 0.03$ is
expected. The visibility to Thomson scattering is defined by
$e^{-\tau_C}$. As long as $\tau_C$ is not too large, ${\cal C}_\ell$
is suppressed by a factor $\exp[-2\tau_C]$ on scales smaller than the
horizon at $z_{\rm reh}$, and in particular over the regime probed by
\cbi, \Boomerang, \DASI, \Maxima, \VSA, etc.

The mechanism for reionization is thought to be the overlap of \ion{H}{2}
regions generated by massive stars housed in the very small, earliest
galaxies to form. Other possibilities, \eg involving particle decays
tuned so that $\tau_C$ would not be too large, are more exotic and
require extra parameters for the theory.  We know that the stars must
form, but little about the efficiency of forming the first
stars. However $z_{\rm reh}$ is necessarily tied to the formation of
nonlinear gas structures, hence to the power spectrum of density
fluctuations.  For the $\Lambda$CDM models preferred by the CMB data,
with $n_s \sim 1$, this implies $z_{\rm reh}$ should not be much more than
15, hence our expectation is that $\tau_C$ should not be much greater
than 0.1. Although the $\tau_C$ likelihoods of
Figs.~\ref{fig:likecbiall},~\ref{fig:likecbidasiB98} now fall off
nicely beyond 0.2 or so, a limit as strong as this still eludes us.
The inability to determine $\tau_C$ with higher precision is
attributable in part to parameter
near-degeneracies~\citep{eb98}. However the fact that we have detected
power at large $\ell$ shows that $z_{\rm reh}$ cannot be too big: \eg
although some pregalactic energy injection at $z \sim 50$ is still
possible, it now seems unlikely that it could have led to full
reionization at such high redshifts.

The late time reheating described above is almost entirely a damping
effect associated with photons coming towards us carrying anisotropy
information being scattered away from our line of sight. There is a
small effect associated with new scattering sources regenerating CMB
anisotropies: the reciprocal effect of photons being scattering into
our line of sight does not appreciably add to the anisotropy we
observe, so the net effect is the damping decline $\propto
e^{-\tau_C}$. The differential visibility ${\cal V}_C \equiv
de^{-\tau_C}/d\ln a = e^{-\tau_C} \bar{n}_e \sigma_T/H(z)$, where
$H(z)$ is the Hubble parameter at redshift $z$ and $a=(1+z)^{-1}$ is
the expansion factor, defines the dominant regions in redshift where
the scattering leads to ``visible'' consequences.  Reionization
results in a bump in ${\cal V}_C$ around $z_{\rm reh}$ with a tail to
lower redshift.

\subsection{Anisotropies from the Recombination Epoch} \label{sec:recombepoch}

We now turn to effects associated with the recombination of the
primordial plasma.  The essential ingredients were worked out
immediately after the discovery of the CMB~\citep{pjep68,ZKS69}. The
novel features are the dominant roles played by the two-photon decay
of the $2s$ state to the $1s$ state and leakage from the Lyman alpha
line associated with the expansion of the Universe. Improvements 
including helium recombination (\eg \citealt{huscottsugwhite95}) and a
more sophisticated treatment of hydrogen recombination were essential
for the high precision era we are entering now \citep{seager99}.

The great simplification afforded by the smallness of the primary
anisotropies is that linear perturbation theory can be used and the
photon transport equations can be decomposed into independent modes
characterized by a comoving wavenumber $k$. Each mode contributes to
${\cal C}_\ell$. However, as mentioned in \S~\ref{sec:intro}, the transport of
the photons through recombination involves all of the complications of
radiative transfer as one passes from an optically thick to an
optically thin ``Thomson scattering atmosphere,'' compounded by a
changing gravitational potential.

Many attempts have been made, in the long history of CMB, to
deconstruct CMB anisotropies into components associated with
baryon-photon acoustic compressions and rarefactions, the Doppler
effect, damping, finite decoupling surface width, polarization
development, and post-decoupling free-streaming propagation. These
have included analytic, semi-analytic, and various small-angle and
large-angle approaches. There were usually two goals: first and
foremost to understand the physics defining the basic features of the
spectra; secondly to make quantitative numerical estimations
appropriate for the computer power of the times. Some used
photon-baryon one-fluid or two-fluid approximations (\eg
\citealt{silk68,weinberg71,pv812fluid,bonom2fluid,vic,doroshtight,starotight,doroshtight2,seljak94,hu94tight,bh95,hu_and_white});
others used these methods in conjunction with other transport
approximations, (\eg Bardeen 1968 (unpublished),
\citealt{pjepyu70,DSZ78,WilsonSilk81,Wilson83,kaiser83pol,be84,VittorioSilk84,be87,jkks96,weinberg01a,weinberg01b,knoxfast02}).

Given the mutual interdependence of the effects, the semi-analytic
methods can only be taken so far, and numerical computation of spectra
is the preferred method for this high precision era of CMB
observations. There were many groups who developed codes to solve the
perturbed Boltzmann--Einstein equations when dark matter was present
prior to and following shortly after the COBE discovery
\citep{be84,be87,VittorioSilk84,VittorioSilk92,eb86,sugiyama1,Gouda91,stompor,cbdes,cdspol,DodelsonJubas,bh95,knox,huscottsugwhite95}.
Most of these solved hierarchies of coupled multipole equations. A
speedy, publicly available and widely used code for evaluation of
anisotropies in a variety of cosmological scenarios, ``CMBfast''
\citep{cmbfast}, using line-of-sight past history integrations of CMB
anisotropy source terms, has come to dominate the scene, used even by
those who developed their own codes and was checked in detail with a
number of these other transport codes. It has also had extensions to
more cosmological models added by a variety of researchers. Another
fast code, ``CAMB'' \citep{camb}, is based on this technique. A
suitably modified CMBfast was used in the construction of the ${\cal
C}_\ell$-database used in \citet{lange01} and in this paper.

\subsection{Numerology of Recombination and Damping} \label{sec:recombdamp}

We have seen that analytic, semi-analytic and deconstruction attempts
have been historically important in quantitative CMB work and
continue to be qualitatively useful in understanding how the various
effects manifest themselves in ${\cal C}_\ell$. Here we shall
concentrate on some of the scales of relevance for this qualitative
description \citep{bh95,hu94tight,hu_and_white,eb98}.

As we take $t_1$ through photon decoupling and recombination, $\tau_C(t_1)$
grows to extreme opaqueness above the redshift $z_{\rm dec}$ where photons
decouple, and where recombination predominantly
occurs.  The differential visibility ${\cal V}_C$ is sharply peaked
for normal recombination and only weakly dependent on cosmological
parameters.  $z_{\rm dec}$ is defined to be where ${\cal V}_C$ has a
peak. A value $\sim 1100$ is obtained for a wide range of
cosmologies. For the $\Omega_{\rm tot}=1$ $\Lambda$CDM cosmologies favored
by our CMB data, $z_{\rm dec} \approx 1050$, with age $t_{\rm dec} \approx
400000 $ yr, (comoving) ``horizon'' scale $c\tau_{\rm dec} \sim 110\, 
\omega_m^{-1/2}\mpc $, and (comoving) distance from us $\chi_{\rm dec
} \approx 0.88 \, (6000\,  \omega_m^{-1/2} \mpc)$. (For Einstein de Sitter
universes, the 0.88 becomes 0.97.) 

The ratio of the density of ordinary matter to relativistic matter at
$z_{\rm dec}$ would then be $\sim 3.3$, and the baryon to photon density
ratio would be $\sim 0.8$. The sound speed at decoupling, $c_s =
(c/\sqrt 3) [1+3\bar{\rho}_B/(4\bar{\rho}_\gamma )]^{-1/2}$ is lowered
over the $c/\sqrt{3}$ for a pure photon gas because of the inertia in
the baryons; for $\omega_b =0.02$, it is $\approx 0.8 c/\sqrt{3}$,
leading to a sound crossing distance $c_s \tau_{\rm dec} \approx 50\,
\omega_m^{-1/2}\mpc $. Since $c_s$ varies with time, an appropriate
average $\bar{c}_s$ should be used, resulting in an adjustment upward
of 12\%. The phase of the waves as they hit the narrow recombination
band, $k\bar{c}_s \tau_{\rm dec}$, determines the oscillations in ${\cal
C}_\ell $, associating peak $m$ with a length scale $\sim (\bar{c}_s
\tau_{\rm dec})/(m\pi)$. 

To convert the comoving distances at $z_{\rm dec}$ to angular scales, we 
divide by $\chi_{\rm dec }$. The component of a wavenumber perpendicular
to the decoupling surface, $k_\perp$ is associated with a multipole
$\ell$, where $k_\perp^{-1} = \chi_{\rm dec} \ell^{-1}$, $\approx 5.3
(1000/\ell)\,  \omega_m^{-1/2} \mpc $ for our $\Lambda$ CDM case. The
mass in matter enclosed within a perturbation of radius $r$ is $M
=2.76 \times 10^{11} \omega_m (4\pi/3) r^3 \msun$. The appropriate top
hat radius of a collapsed object that forms from waves associated with
a band about $k$ is $r \sim 2k^{-1}$ \citep{bond_and_myers}. This
gives a rich cluster mass, $1.3 \times 10^{15}\, 
\omega_m^{-1/2} (1000/\ell)^3 \msun$, for $\ell \sim 1500$.  

Converting peaks in $k$-space into peaks in $\ell$-space is obscured
by projection effects and the influence of other sources such as the
Doppler term. The conversion of the oscillations into peak locations
in ${ \cal C}_\ell$ gives $\ell_{pk,m} \sim f_m m \pi \chi_{\rm
dec}/(\bar{c}_s \tau_{\rm dec})$, where the numerically estimated
$f_m$ factor is $\approx 0.75$ for the first peak, approaching unity
for higher ones. These numbers accord reasonably well with the values
we obtain when we average over the probability distribution for our
${\cal C}_\ell$-database. Using all-data, and the flat+wk-$h$+LSS
prior, the first five peak locations are at $221 \pm 2$, $536 \pm 5$,
$814 \pm 17$, $1129 \pm 18$, $1427 \pm 20$. The interleaving dips are
at $413 \pm 4$, $673 \pm 8$, $1015 \pm 10$, $1310 \pm 14$, similar to
the predicted $(m+1/2)$ spacing. For all-data and the wk-$h$ prior,
the positions are very similar, but as expected the errors are
slightly larger. We saw in Paper III that the ``model-independent''
estimation of peak positions from the data accord reasonably well with
these ${\cal C}_\ell$-database determinations.

The electron density falls dramatically through decoupling: the local
power law index, $p=-d\ln n_e /d\ln a$, rises from its low and high
$z$ asymptotic values of 3 to a maximum of about 15, with 10 to 14
typical at $z_{\rm dec}$, the range depending upon the model; \eg for
$\Lambda$CDM, $p \approx 12$. A ``Gaussian'' width of decoupling in
$\ln a/a_{\rm dec}$ can be estimated analytically from $p$:
$\sigma_{a,\rm dec} \approx (p-1)^{-1}$ (apart from a small correction
factor associated with the change of $p$). For $\Lambda$CDM, $
\sigma_{a,\rm dec} \approx 0.08$. (Although the distribution is somewhat
skewed, a Gaussian fit to ${\cal V}_C$ turns out to be a reasonably
good approximation over the dominant range, and estimating
$\sigma_{a,\rm dec}$ from the FWHM of exactly ${\cal V}_C$ yields
0.06 to 0.1 in good accord with the analytic estimates.)  The
corresponding comoving length scale, $R_{C,dec} \approx \sigma_{a,\rm dec}
[H(z_{\rm dec})a_{\rm dec}]^{-1}$, is $\sim 7\,  \omega_m^{-1/2}\mpc $ for
$\Lambda$CDM, to be compared with $c_s \tau_{\rm dec}$. Because this is
parallel to the line of sight, $R_{C,\rm dec}$ does not project onto an
angular scale we can observe, but perpendicular components of this 
characteristic size would have $\ell \sim 820$. 

 A combination of viscous (Silk) damping and fuzziness damping
diminishes the amplitude of the acoustic peaks. The two effects occur
simultaneously, intertwined by the complexities of the transport, but
are estimated differently. 

Earlier than decoupling, the photons and baryons are so tightly
coupled by Thomson scattering that they can be treated as a single
fluid with sound speed $c_s$, shear viscosity $({4}/ (15f_\eta))
\bar{\rho}_\gamma /(n_e\sigma_T)$, zero bulk viscosity, and thermal
conductivity $\kappa_{\gamma} = (4\rho_\gamma /(3
T_\gamma))/(n_e\sigma_T)$. Here $f_\eta$ is $3/4$ if Thomson
scattering is fully treated and 1 if polarization and the angular
dependence of the Thomson cross section are ignored. Silk damping has
usually been estimated using a WKB approximation to these one-fluid 
equations, which results in an overall damping multiplier of
form $e^{-(k\sigma_{D} \tau_{\rm dec})^2/2}$ for $z > z_{\rm dec}$ multiplying
the $kc_s\tau_{\rm dec}$ terms which give the acoustic oscillations. The
parameter $\sigma_D$ is an integral of the damping rate involving the
shear viscosity and thermal conductivity, 
\begin{eqnarray}
\sigma_D^2 &\approx& {1\over (p-1)(p -1/2)(15f_\eta )(1+R)} \nonumber\\
           &       & + {R^2\over (p-1)(p -1/2) 12 (1+R)^2},  \nonumber\\
 R &\equiv& {3 \bar{\rho}_B
\over 4\bar{\rho}_\gamma} . \label{eq:sigmaD}
\end{eqnarray} 
With $f_\eta =3/4$ and $10\lta p \lta 14$, we have $0.02 \lta \sigma_D
\lta 0.03$ over a wide range of cosmological parameters, 0.023 for
$\Lambda$CDM, with only a weak sensitivity to $\omega_b$. The first
term is from shear viscosity, the second is from thermal
conductivity. For $\Lambda$CDM with $\omega_b=0.02$, the ratio is 5 to
1. Although polarization increases $\sigma_D$ by 10\%, we cannot
determine the damping scale with such accuracy with the CBI data.

For this $\sigma_D$, we get $\sigma_D c\tau_{\rm dec} \approx 2.4\,
\omega_m^{-1/2} \mpc$, giving a scale $\ell_D \sim 2160$.  However,
the tight coupling equations break down as the radiation passes
through decoupling, so it is better to treat $\sigma_D$ as a
phenomenological factor and match it to numerical results. An estimate
of the Silk damping scale (given as well by \citealt{bh95}) was
$k_{\rm Silk}^{-1} \approx 3.8\, \omega_{m}^{-1/2} \mpc$, with angular
size $2.5^{\prime}$ and $\ell_{\rm Silk} \approx 1390$. A more
sophisticated phenomenology of numerical ${\cal C}_\ell$-results
adopted damping envelope functions, $\exp[-(\ell /\ell_D)^{m_D}]$,
multiplying ``undamped ${\cal C}_\ell$'s'' and provided fitting
formulas for $\ell_D$ and $m_D$ \citep{hu_and_white}. For the
$\Lambda$CDM parameters used here, we get $\ell_D =1345$, with a power
$m_D=1.26$, in good accord with the $\ell_{\rm Silk}$ estimate. Note
that the falloff is not as steep as the WKB Gaussian would
predict. When we average $\ell_D$ over the database, we get $1352 \pm
15$ for all-data and the flat+wk-$h$+LSS prior. 

The fuzziness damping acts only on $k_\parallel$, the component of the
waves through the decoupling surface: destructive interference from
both peaks and troughs occurs for waves with $k_\parallel R_{C,dec} >
\pi$, but there is none if the photons are only received from either
peaks or troughs, but not both, the case if oscillations are along the
surface, or if the wavenumbers are small. (The WKB tight-coupling
solution does in fact calculate a version of fuzziness damping along
with other transport effects, but the $k_\perp$--$k_\parallel$
asymmetry is obscured by the truncation of the $\ell$-hierarchy at
such low $\ell$: up to $\approx \tau_{\rm dec}$, higher moments are
strongly damped, but this is not correct as the photons pass through
$z_{\rm dec}$.) In the Gaussian approximation to ${\cal V}_C$, fuzziness
damping acts on $\Delta T/T$ through a multiplier $e^{-(k_\parallel
R_{C,dec} )^2/2}$. Because it acts asymmetrically, it is not as
dramatic a drop as in the WKB case even though $R_{C,dec}$ is
bigger than $k_{\rm Silk}^{-1}$. A simple estimate of a fuzziness damping
scale angle-averages $(k_\parallel R_{C,dec})^2$, reducing the
effective filter to $R_{C,dec}/\sqrt{3}$, giving $\ell_{C,dec} \sim
1420$, similar to $\ell_{\rm Silk}$ and $\ell_D$. 

If we use $\ell_D =1345$ and the damping envelope to ``correct'' the
heights of the peaks and dips determined along with their $\ell$-space
locations, the peak power bounces between 4000 and 2500 $\mu\rm K^2$
(except for the first peak), and the dip power between 2600 and 1400,
``correcting'' for a significant fraction of the factor of about five
raw variations.  It is certainly an attractive proposition to directly
translate the CMB data into accurate determination of the physical
scales operating at decoupling. However, the intertwining of transport
effects makes the use of a parameterized ${\cal C}_\ell$ model space a
more robust proposition.


\section{Conclusions} \label{sec:concl}

The \cbi\ provides a unique view of the CMB spectrum extending to much
higher $\ell$ than previous experiments which have detected primary
anisotropies, and well into the multipole region of the spectrum
dominated by the damping of fluctuations at decoupling due to
viscosity in the photon-baryon fluid and the finite thickness of the
last scattering region.  The CBI observations indicate a flat universe
with a scale-invariant primordial fluctuation spectrum consistent with
the inflationary model; and in addition they indicate a low matter
density, a baryon fraction consistent with Big Bang Nucleosynthesis, a
non-zero cosmological constant, and a cosmological age consistent with
the ages of the oldest stars in globular clusters.

These results hold for the whole CBI data set, and in addition they
hold for a subset of the data restricted to $610<\ell<3500$.  These
findings are therefore independent of the spectrum over the
$\ell$-range of the first and second acoustic peaks, and thus provide
{\it independent} confirmation of the major results determined by
other CMB experiments from observations that span the first two or
three acoustic peaks.  This independent confirmation of the major
results gives much confidence that the key assumptions of minimal
inflationary models are correct, especially the assumption that the
primordial fluctuation spectrum does not have significant fine
structure.  If there were significant fine structure, it is extremely
unlikely that spectral studies over different $\ell$-ranges would
yield the same values of key cosmological parameters.  
The good agreement with the results from lower $\ell$ also
demonstrates that the recombination theory of the simple model is
substantially correct.

In more detail: the CBI observations from the first year of observing,
when combined with DMR, gives the following key cosmological results,
as discussed in \S~\ref{sec:cbionly}.  Under the weak-$h$+LSS priors
we find $\Omega_{\rm tot} = 1.05_{-0.08}^{+0.08}$, and
$n_s = 1.07_{-0.10}^{+0.13}$, consistent with inflationary models;
$\Omega_{\rm cdm}h^2=0.10_{-0.03}^{+0.04}$, and, in addition, identifying the
excess energy density with the cosmological constant, we find
$\Omega_{\Lambda} = 0.67_{-0.13}^{+0.10}$.  When the more restrictive
priors, flat+weak-$h$+LSS, are used, we find $\Omega_{\rm
cdm}h^2 = 0.11_{-0.02}^{+0.02}$, consistent with large scale structure
studies; $\Omega_b h^2 = 0.024_{-0.009}^{+0.011}$, consistent with Big
Bang Nucleosynthesis; $\Omega_m = 0.34\pm 0.12$, and
$\Omega_b = 0.057\pm 0.020$, indicating a low matter density universe;
$h = 0.66_{-0.11}^{+0.11}$, consistent with the recent determinations
of the Hubble Constant based on the recently revised Cepheid
period-luminosity law; and $t_0=14.2_{-1.3}^{+1.3}$ Gyr, consistent
with cosmological age estimates based on the oldest stars in globular
clusters.

These values for key cosmological parameters are in remarkably good
agreement with those determined in other recent CMB experiments.  As
pointed out above, this is highly significant, since the \cbi\
cosmology is based on a higher $\ell$-range than has previously been
used and leverages principally off the damping tail region of the
spectrum rather than the first acoustic peak.

Another unique aspect of the 500--3500 $\ell$-range that \cbi\ has
probed is that the angular scales correspond to the 3D wavenumbers of
structures that collapsed to produce clusters of galaxies, ranging
from those with masses as low as $10^{14} M_\sun$, 
to large
superclusters with masses $\sim 10^{16} M_\sun$.  Hence the CBI
observations span the whole range of masses from small groups of
galaxies to large superclusters.  Thus detection of CMB power in this
region provides a direct link between the small $\Delta T$
fluctuations at the time of photon decoupling and the nonlinear
density amplitudes on those scales today. This provides further strong
support for the gravitational instability picture of structure
formation.

The \cbi, \Boomerang, \DASI, \Maxima, and \VSA\ observations are consistent
with one another over the entire range of overlapping coverage in
$\ell$.  The consistency between the five data sets, obtained by
different experiments using different observation strategies on
different parts of the sky, eliminates many sources of systematic
error as a potential cause for concern. Figure~\ref{fig:2sigbothall}
shows concordance of the different experiments in our
minimal-inflation parameter space.

Given the small signal levels that are being studied here, it is
remarkable that the agreement between the recent CMB experiments
(\TOCO, \Boomerang, CBI, \DASI, \Maxima, and \VSA) should be so good.  The
variety of techniques employed, combined with the extension of the
spectrum to high-$\ell$ provided by the CBI data and the high
$\ell$-resolution data at low multipoles provided by the \Boomerang,
\DASI, \Maxima, and \VSA\ experiments, makes for a compelling case that both
the observations and the cosmological results are robust.

We have not treated several other parameters that could be of
relevance to the inflation-based model. In place of $\Lambda$CDM,
$Q$CDM is receiving much attention, with $Q$ an ultra-low mass scalar
field, often called quintessence, that dominates at late times. Thus
$\Omega_Q$ replaces $\Omega_\Lambda$ and an effective $Q$-dynamics is
cast in terms of a mean pressure-to-density ratio $w_Q =
\bar{p}_Q/\bar{\rho}_Q$, an effective equation of state (EOS). (The
dynamics of $Q$ is more complex than this, since $Q$ is expected to be
spatially as well as temporally varying. There is also no good
candidate for a theory of $Q$.)  For $\Lambda$, $w_Q=-1$, but $w_Q < -
1/3$ would get our patch of the Universe into acceleration. $w_Q$ is
not well determined by CMB data and we require supernova information
to get a useful constraint on it \citep[e.g.,][]{BondQ2000,bmrst02}. The CMB by
itself also is insensitive to the addition of a light massive
neutrino (H$\Lambda$CDM models) since there is only a small effect on
${\cal C}_\ell$. LSS can add discriminatory power but it simply shifts
the result to slightly lower (but still nonzero) $\Omega_\Lambda$
\citep[e.g.,][]{Bondnu2000,Pogosyan95}.

The influence of a possible gravity-wave component will be explored
elsewhere but the main result is that it is also  expected to have
little effect on the main cosmological parameters presented here. It
certainly would have no impact on the angular scales probed by \cbi.

Although our results show a tantalizing drop in the $\tau_C$
likelihood beyond 0.2, as described in \S~\ref{sec:damp},there is some
distance to go to get a detection at the $\sim 0.1$ values simple
theoretical predictions give for $\Lambda$CDM. Significant early
energy injection in the medium cannot have occurred. This constrains a
number of possible, if not probable, scenarios: it is always possible
to generate many ionizing stars early by amplifying structure
formation from rare collapses that occurred at high redshift, having a
non-Gaussian component at small masses, or making the primordial
spectrum bluer on small scales.

The dominant feature of the \cbi\ data is the overall decline in the
power with increasing $\ell$, a strong prediction of the basic theory
of photon decoupling, with a damping scale now moderately well
determined, as described in \S~\ref{sec:damp}. The way the simplest
inflation-based models have survived the dramatic extension to higher
$\ell$ over what previous experiments probed is rather amazing. As
well \cbi\ has provided further evidence that the peaks and dips
associated with acoustic oscillations continue to higher $\ell$ and
are in roughly the right locations given by the emerging $\Lambda$CDM
concordance model. Thus, the CMB results from the CBI and from the
\Boomerang, \DASI, \Maxima, and \VSA\ experiments, both in isolation and
particularly in combination, strongly support the chief predictions of
the inflation paradigm --- the geometry of the universe is flat, the
initial density perturbations are nearly scale-invariant, and the
density of mass-energy in the universe is dominated by a form other
than ordinary matter. Simple models in which structure formation is
driven by topological defects are difficult to reconcile with the CMB
observations. These conclusions are considerably strengthened by the
inclusion of other cosmological results such as measurements of the
Hubble constant, the amplitude and shape of the power spectrum, and
the accelerating expansion rate. These all point towards a nonzero
cosmological constant.

\acknowledgements We thank Roger Blandford, Marc Kamionkowski, and
Sterl Phinney for useful discussions.  We gratefully acknowledge the
generous support of Maxine and Ronald Linde, Cecil and Sally
Drinkward, Barbara and Stanley Rawn, Jr., and the Kavli Institute, and the
strong support of the provost and president of the California
Institute of Technology, the PMA division chairman, the director of
the Owens Valley Radio Observatory, and our colleagues in the PMA
Division.  This work was supported by the National Science Foundation
under grants AST 94-13935, AST 98-02989, and AST 00-98734.  Research
in Canada is supported by NSERC and the Canadian Institute for
Advanced Research. The computational facilities at Toronto are funded
by the Canadian Fund for Innovation. 
LB and JM acknowledge support from the Chilean Center for Astrophysics FONDAP No.~15010003.
We thank CONICYT for granting permission to
operate within the Chanjnantor Scientific Preserve in Chile.

\appendix \section{Testing the Offset Lognormal Approximation to the
Likelihood }
\label{app:ps}

\begin{figure}
\epsscale{0.6}
\plotone{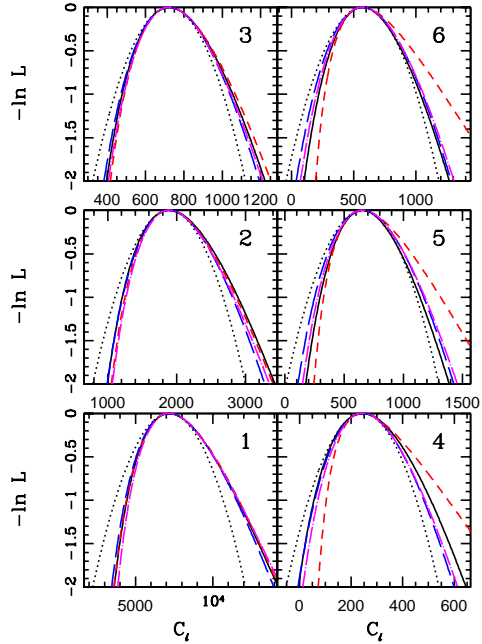}
\caption{This shows how well the offset lognormal approximation
(long-dashed dark blue) does in reproducing the likelihood functions
(solid black) for individual bandpowers $q^B$ when the rest of them
are fixed at their maximum likelihood values. This is for the first 6
of the 7 \dCBI\ bands. The offset logormal approximation with $q_{\rm
Nt}^B$ determined from our \cbi-pipeline reproduces the likelihood
function quite well to beyond 2-$\sigma$. Gaussian (dotted black) and
log-normal (short-dashed red) distributions are shown for comparison, the
former a better fit in the noise-dominated bands, the latter in the
cosmic variance dominated bands. The offset log-normal interpolates
nicely between the two regimes. The equal-variance approximation
(dash-dotted magenta) \citep{BJK2000} also fits quite well. This result is a
cornerstone of parameter estimation, whether it be for optimal
spectrum combinations of parameterized ${\cal C}_\ell$ shapes or for
cosmological parameter estimations in a ${\cal C}_\ell$-database.  }
\label{fig:offsetlognormpfdeep}
\end{figure}

\begin{figure}
\epsscale{0.6}
\plotone{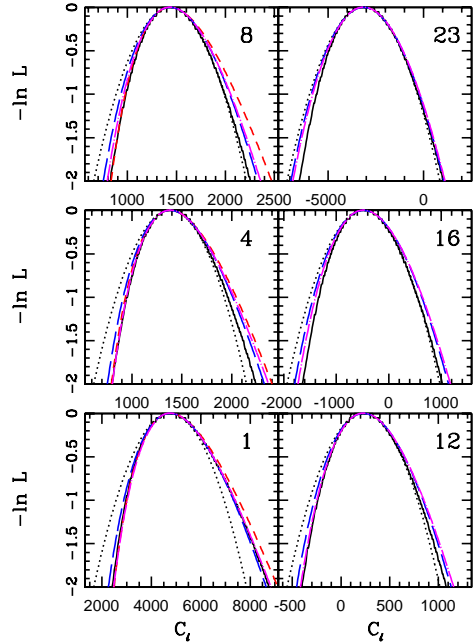}
\caption{Same as Fig.~\ref{fig:offsetlognormpfdeep}, but for a
selected spread of \mCBIofg\ bands showing again that the offset
lognormal approximation works quite well for the \cbi\ mosaic
data. Another necessary ingredient is that the bands must not be so
closely spaced that the band-to-band correlation as estimated by the
inverse Fisher matrix is strong since a weak approximation is used for
this band-to-band coupling. Both the $\Delta \ell =140$ and $\Delta\ell=200$
binnings are adequate choices, and deliver similar parameter
determinations. }
\label{fig:offsetlognormpfmos}
\end{figure}

Whether we are interested in estimating parameters which are
cosmological or those defining an optimal spectrum, the full 
likelihood surface is required, not just its
Gaussian approximation which is valid only in the immediate
neighborhood of the maximum. This can  in principle 
only be done by full calculation, though  in practice 
there are two analytic approximations that have have been shown to fit
the one-point distributions quite well in all the cases tried
\citep{BJK2000,Netterfield02}. We show these work well for the \cbi\
deep and mosaic cases too in Figures~\ref{fig:offsetlognormpfdeep}
and~\ref{fig:offsetlognormpfmos}. The simplest and most often used
analytic form is the ``offset lognormal'' distribution. This is
obtained by taking a Gaussian in the variable $z^B=\ln (q^B
+q_{Nt}^B)$, where the offset $q_{Nt}^B$ is the effective noise in the
experiment:
\begin{eqnarray}
&&{\cal P}(q) \propto \exp[-\half \sum_{BB^\prime} (z-\bar{z})^B
{\cal F}^{(z)}_{BB^\prime} (z-\bar{z})^{B^\prime}], \nonumber \\
&& {\cal
F}^{(z)}_{BB^\prime} =(\bar{q}^B+q_{\rm Nt}^B) {\cal
F}^{(q)}_{BB^\prime}(\bar{q}^{B^\prime}+q_{\rm Nt}^{B^\prime}). \label{eq:xb}
\end{eqnarray}
Here ${\cal F}^{(q)}_{BB^\prime}$ is the curvature matrix for the
bandpowers $q^B$ and ${\cal F}^{(z)}_{BB^\prime}$ is its local
transformation to the $z^B$ variables. We sometimes use the ensemble
average value of ${\cal F}^{(q)}_{BB^\prime}$, which is the Fisher
matrix, rather than the curvature matrix.  Other approaches for
evaluating $q_{Nt}^B$ are reviewed by \citet{bc01}.
Figures~\ref{fig:offsetlognormpfdeep} and ~\ref{fig:offsetlognormpfmos}
show how the offset lognormal approximation provides an accurate
description of the likelihood in individual bands for both deep and
mosaic data to beyond the $2\sigma$ level. It is also evident that the
Gaussian and pure lognormal approximations  do not fit
the likelihoods as well, the former working best in low
signal-to-noise bands, the latter in high signal-to-noise bands. 

To compare a given theory with spectrum ${\cal C}_{T\ell}(y^a)$ with
the data using equation~(\ref{eq:xb}), the  model  $q^B$'s need
to be evaluated with a specific choice for $\varphi_{B\ell}$:
\begin{equation}
q_B = {\cal I} [ {\cal C}_\ell \varphi_{B\ell}]/{\cal I} [ {\cal
C}^{(s)}_\ell \varphi_{B\ell}],
 \ {\rm where} \ 
{\cal I} [f_\ell]
\equiv \sum_\ell f_\ell {\ell +\half \over \ell (\ell +1 )}     
\label{eq:winfunc}
\end{equation}
is the discrete ``logarithmic integral'' of a function $f_\ell$. The
associated bandpower in $(\mu {\rm K})^2$ is
\begin{equation}
{\cal C}_B =  q^B {\cal C}^{(s)}_B  \, ({\rm no\, sum}), 
\ {\cal C}^{(s)}_B \equiv {\cal I} [ {\cal
C}^{(s)}_\ell \varphi_{B\ell}]/{\cal I} [ \varphi_{B\ell}]\, .
\end{equation}
As discussed by \citet{BJK2000}, \citet{knox00}, and \citet{bc01}, the window function
$\varphi_{B\ell} $ used for this purpose (and therefore with a
different notation than the $\psi_{b\ell}$ defined above) is somewhat
arbitrary. The simplest choice is again that of a top hat $\chi_b
(\ell )$. We prefer to use the signal-to-noise windows $W_{B}(\ell)$
derived from the experiment (Paper~IV). We have also considered the
truncated form $W_{B}(\ell )\chi_{B\ell}$. We find that the
cosmological parameters we derive in \S~\ref{sec:cbionly} and
\S~\ref{sec:allparams} are insensitive to which form we use. (For
example, the largest change occurs in $\omega_b$, by 4\% in the mean
and 10\% in the error. This is also evident from
Figure~\ref{fig:compare3mos}, which uses $\varphi_{B\ell}=W_{B\ell}$ for
the data to create an optimal spectrum with
$\psi_{B\ell}=\chi_{B\ell}$; the results are almost identical to
the original data.)

If we decrease the width $\Delta \ell$ of bins, we develop more
correlation between neighboring bands. The offset lognormal
approximation  has only been shown to give  a good fit
when considering individual bins. We therefore  require 
weak band-to-band coupling. We have carried out extensive tests on the
dependence of the parameter determinations on binning width and
positioning.   We showed in \S~\ref{sec:cbionly}  that the inferred
cosmologies are insensitive to the choice of bin boundaries.

\end{document}